\documentclass[preprint2,numberedappendix,tighten,twocolappendix]{aastex6} 
\shorttitle{Spherical Fourier-Bessel Intensity Mapping Analysis}
\shortauthors{Liu et al.}

\usepackage{amsmath}
\usepackage{graphicx}
\usepackage[figuresright]{rotating}
\usepackage{natbib}
\usepackage{ctable}
\newcommand{\rhat}{\hat{\mathbf{r}}}
\newcommand{\khat}{\hat{\mathbf{k}}}
\newcommand{\btheta}{\boldsymbol \theta}
\newcommand{\appropto}{\mathrel{\vcenter{
  \offinterlineskip\halign{\hfil$##$\cr
    \propto\cr\noalign{\kern2pt}\sim\cr\noalign{\kern-2pt}}}}}

\begin{document}
\title{Spherical Harmonic Analyses of Intensity Mapping Power Spectra}

\author{
Adrian Liu$^{\dagger}$,
Yunfan Zhang,
Aaron R. Parsons
}
\affil{Department of Astronomy and Radio Astronomy Laboratory, University of California Berkeley, Berkeley, CA 94720, USA}
\email{acliu@berkeley.edu}
\begin{abstract}
Intensity mapping is a promising technique for surveying the large scale structure of our Universe from $z=0$ to $z \sim 150$, using the brightness temperature field of spectral lines to directly observe previously unexplored portions of out cosmic timeline. Examples of targeted lines include the $21\,\textrm{cm}$ hyperfine transition of neutral hydrogen, rotational lines of carbon monoxide, and fine structure lines of singly ionized carbon. Recent efforts have focused on detections of the power spectrum of spatial fluctuations, but have been hindered by systematics such as foreground contamination. This has motivated the decomposition of data into Fourier modes perpendicular and parallel to the line-of-sight, which has been shown to be a particularly powerful way to diagnose systematics. However, such a method is well-defined only in the limit of a narrow-field, flat-sky approximation. This limits the sensitivity of intensity mapping experiments, as it means that wide surveys must be separately analyzed as a patchwork of smaller fields. In this paper, we develop a framework for analyzing intensity mapping data in a spherical Fourier-Bessel basis, which incorporates curved sky effects without difficulty. We use our framework to generalize a number of techniques in intensity mapping data analysis from the flat sky to the curved sky. These include visibility-based estimators for the power spectrum, treatments of interloper lines, and the ``foreground wedge" signature of spectrally smooth foregrounds.
\end{abstract}

\section{Introduction}
\label{sec:Intro}
{\let\thefootnote\relax\footnote{$^{\dagger}$Hubble Fellow}}
\setcounter{footnote}{0}
In recent years, intensity mapping has been hailed as a promising method for conducting cosmological surveys of unprecedented volume. In an intensity mapping survey, the brightness temperature of an optically thin spectral line is mapped over a three-dimensional volume, with radial distance information provided by the observed frequency (and thus redshift) of the line. By observing brightness temperature fluctuations on cosmologically relevant scales (without resolving individual sources responsible for the emission or absorption), intensity mapping provides a relatively cheap way to survey our Universe. In addition, with an appropriate choice of spectral line and a suitably designed instrument, the volume accessible to an intensity mapping survey is enormous. This allows measurements to be made over a large number of independent cosmological modes, providing highly precise constraints on both astrophysical and cosmological models. For example, intensity mapping experiments tracing the $21\,\textrm{cm}$ hyperfine transition of hydrogen can easily access $\sim 10^9$ independent modes, which is much greater than the $\sim 10^6$ accessible to the Cosmic Microwave Background, in principle unlocking a far greater portion of the available information in our observable Universe \citep{loeb_and_zaldarriaga2004,mao_et_al2008,tegmark_and_zaldarriaga2009,ma_and_scott2016,scott_et_al2016}.

A large number of intensity mapping experiments are in operation, and more have been proposed. Post-reionization neutral hydrogen $21\,\textrm{cm}$ intensity mapping is being conducted by the Canadian Hydrogen Intensity Mapping Experiment \citep{bandura_et_al2014}, the Green Bank Telescope \citep{masui_et_al2013}, Tianlai telescope \citep{chen_et_al2012}, Baryon Acoustic Oscillations from Integrated Neutral Gas Observations project \citep{battye_et_al2013}, Hydrogen Intensity and Real-time Analysis eXperiment \citep{newburgh_et_al2016}, and BAORadio \citep{ansari_et_al2012}. These experiments use neutral hydrogen as a tracer of the large scale density field, with a primary scientific goal of constraining dark energy via measurements of the baryon acoustic oscillation feature from $0 < z < 4$ \citep{wyithe_et_al2008,chang_et_al2008,pober_et_al2013a}. At $z \sim 2$ to $3.5$, data from the Sloan Digital Sky Survey have been used for Ly $\alpha$ intensity mapping \citep{croft_et_al2016}. Other experiments such as the CO Power Spectrum Survey \citep{keating_et_al2015,keating_et_al2016} and the CO Mapping Array Pathfinder \citep{li_et_al2016} use CO as a tracer of molecular gas in the epoch of galaxy formation at roughly $z \sim 2$ to $3$. Using [CII] instead is the Spectroscopic Terahertz Airborne Receiver for Far-InfraRed Exploration (operating at $0.5 < z < 1.5$; \citealt{uzgil_et_al2014}), and the Tomographic Ionized carbon Mapping Experiment (operating at $5 < z < 9$; \citealt{crites_et_al2014}). The highest redshift bins of the latter encroach upon the Epoch of Reionization (EoR), when the first galaxies systematically reionized the hydrogen content of the intergalactic medium. Extending into the EoR, intensity mapping efforts are mainly focused around the $21\,\textrm{cm}$ line. The Donald C. Backer Precision Array for Probing the Epoch of Reionzation array (PAPER; \citealt{parsons_et_al2010}), the Low Frequency Array \citep{van_haarlem_et_al2013}, the Murchison Widefield Array \citep{bowman_et_al2012,tingay_et_al2013}, the Giant Metrewave Radio Telescope \citep{paciga_et_al2013}, the Long Wavelength Array (M. W. Eastwood et al., in prep.), 21 Centimeter Array \citep{huang_et_al2016,zheng_et_al2016}, and the Hydrogen Epoch of Reionization Array \citep{deboer_et_al2016} are radio interferometers that aim to use the $21\,\textrm{cm}$ line to probe the density, ionization state, and temperature of hydrogen in the range $6 < z < 13$ and beyond. The future Square Kilometre Array \citep{mellema_et_al2015} will provide yet more collecting area for $21\,\textrm{cm}$ intensity mapping to complement the aforementioned experiments. With such a large suite of instruments covering an expansive range in redshift, tremendous opportunities exist for understanding the formation of the first stars and galaxies via direct measurements of the IGM during all the relevant epochs \citep{hogan_and_rees1979,scott_and_rees1990,madau_et_al1997,tozzi_et_al2000}, as well as fundamental cosmological parameters \citep{mcquinn_et_al2006,mao_et_al2008,visbal_et_al2009,clesse_et_al2012,liu_et_al2016} and exotic phenomena such as dark matter annihilations \citep{valdes_et_al2013,evoli_et_al2014}.

Despite its promise, intensity mapping is challenging, and to date the only positive detections have been tentative detections of Ly $\alpha$ at $z \sim 2$ to $3.5$ \citep{croft_et_al2016} and CO from $z\sim 2.3$ to $3.3$ \citep{keating_et_al2016}, as well as detections of HI at $z\sim 0.8$ via cross-correlation with optical galaxies \citep{chang_et_al2010,masui_et_al2013}. To realize the full potential of intensity mapping, it is necessary to overcome a large number of systematics. A prime example would be radiation from foreground astrophysical sources, which are particularly troublesome for HI intensity mapping. Especially at high redshifts, foregrounds add contaminant emission to the measurement that are orders of magnitude brighter than the cosmological signal \citep{dimatteo_et_al2002,santos_et_al2005,wang_et_al2006,deOliveiraCosta_et_al2008,sims_et_al2016}. Low frequency measurements (for instance, those targeting the $21\,\textrm{cm}$ EoR signal), are mainly contaminated by broadband foregrounds such as Galactic synchrotron emission or extragalactic point sources (whether they are bright and resolved or are part of a dim and unresolved continuum). These foregrounds are typically less dominant at the higher frequencies and are thus easier (though still challening) to handle for CO or [CII] intensity mapping experiments. However, such experiments must also contend with the problem of interloper lines, where two spectral lines of different rest wavelengths may redshift into the same observation band, leading to confusion as to which spectral line has been observed.

In addition to astrophysical foregrounds, instrumental systematics must be well-controlled for a successful measurement of the cosmological signal. Among others, these systematics include beam-forming errors \citep{neben_et_al2016b}, radio frequency interference \citep{offringa_et_al2013,offringa_et_al2015,huang_et_al2016}, polarization leakage \citep{geil_et_al2011,moore_et_al2013,shaw_et_al2014b,sutinjo_et_al2015,asad_et_al2015,moore_et_al2015,kohn_et_al2016}, calibration errors \citep{newburgh_et_al2014,trott_and_wayth2016,barry_et_al2016,patil_et_al2016}, and instrumental reflections \citep{neben_et_al2016a,ewall-wice_et_al2016a,thyagarajan_et_al2016}.

In this paper, we focus specifically on measurements of the power spectrum $P(k)$ of spatial fluctuations in brightness temperature, where roughly speaking, the temperature field is Fourier transformed and then squared. In diagnosing the aforementioned systematics as they pertain to spatial fluctuation experiments, it is helpful to decompose the fluctuations into modes that separate purely angular fluctuations from purely radial fluctuations from those that are a mixture of both.
In recent years, for example, simulations and measured upper limits of the $21\,\textrm{cm}$ power spectrum have often been expressed as cylindrically binned power spectra. To form cylindrically binned power spectra,  one begins with unbinned power spectra $P(\mathbf{k})$, where $\mathbf{k}$ is the three-dimensional wavevector of spatial Fourier modes. If the field of view is narrow, there exists a particular direction that can be identified as the line-of-sight (or radial) direction. One of the three components of $\mathbf{k}$ can then be chosen to lie along this direction and labeled $k_\parallel$ as a reminder that it is \emph{parallel} to the line-of-sight. The remaining two components---which we arbitrarily designate $k_x$ and $k_y$ in this paper---describe transverse (i.e., angular fluctuations), and have a magnitude $k_\perp \equiv \sqrt{k_x^2 + k_y^2}$. Binning $P(\mathbf{k})$ along contours of constant $k_\perp$ gives $P(k_\perp, k_\parallel)$, the cylindrically binned power spectrum.

Expressing the power spectrum as a function of $k_\perp$ and $k_\parallel$ is a powerful diagnostic exercise because intensity mapping surveys probe line-of-sight fluctuations in a fundamentally different way than the way they probe angular fluctuations. Systematics are therefore usually anisotropic and have distinct signatures on the $k_\perp$-$k_\parallel$ plane \citep{morales_and_hewitt2004}. For example, cable reflections and bandpass calibration errors tend to appear as features parallel to the $k_\parallel$ axis \citep{dillon_et_al2015,ewall-wice_et_al2016b,jacobs_et_al2016}. Thus, the cylindrically binned power spectrum is a useful intermediate quantity to compute before one performs a final binning along constant $k \equiv \sqrt{k_\perp^2 + k_\parallel^2}$ to give an isotropic power spectrum $P(k)$.

The diagnostic capability of $P(k_\perp, k_\parallel)$ is particularly apparent when considering foregrounds. Assuming that they are spectrally smooth, foregrounds preferentially contaminate low $k_\parallel$ modes, since $k_\parallel$ is the Fourier conjugate to line-of-sight distance, which is probed by the frequency spectrum in intensity mapping experiments. The situation is more complicated for the (large) subset of intensity mapping measurements that are performed on interferometers. Interferometers are inherently chromatic in nature, causing intrinsically smooth spectrum foregrounds to acquire spectral structure, which results in leakage to higher $k_\parallel$ modes. Even this leakage, however, has been shown in recent years to have a predictable ``wedge" signature on the $k_\perp$-$k_\parallel$ plane, limiting the contaminated region to a triangular-shaped region at high $k_\perp$ and low $k_\parallel$ \citep{Datta2010,Vedantham2012,Morales2012,Parsons_et_al2012b,Trott2012,Thyagarajan2013,pober_et_al2013b,dillon_et_al2014,Hazelton2013,Thyagarajan_et_al2015a,Thyagarajan_et_al2015b,liu_et_al2014a,liu_et_al2014b,chapman_et_al2016,pober_et_al2016,seo_and_hirata2016,jensen_et_al2016,kohn_et_al2016}. In fact, the foreground wedge is considered sufficiently robust that some instruments have been designed around it \citep{pober_et_al2014,deboer_et_al2016,dillon_et_al2016,neben_et_al2016a,ewall-wice_et_al2016a,thyagarajan_et_al2016}, implicitly pursuing a strategy of foreground avoidance where the power spectrum can be measured in relatively uncontaminated Fourier modes outside the wedge. This mitigates the need for extremely detailed models of the foregrounds, providing a conservative path towards early detections of the power spectrum.

Despite its utility, the $k_\perp$-$k_\parallel$ power spectrum is limited in that it is ultimately a quantity that is only well-defined in the flat-sky, narrow field-of-view limit, where a single line-of-sight direction can be unambiguously defined. For surveys with wide fields of view, different portions of the survey have different lines of sight that point in different directions with respect to a cosmological reference frame. Note that this is a separate problem from that of wide-field imaging: even if one's imaging software does not make any flat-sky approximations (so that the resulting images of emission within the survey volume are undistorted by any wide-field effects), the act of forming a power spectrum on a $k_\perp$-$k_\parallel$ invokes a narrow-field approximation. If one insists on forming $P(k_\perp, k_\parallel)$ as a diagnostic, the simplest way to do so is to split up the survey into multiple small patches that are individually small enough to warrant a narrow-field assumption. A separate power spectrum can then be formed from each patch by squaring the Fourier mode amplitudes, and the resulting collection of power spectra can then be averaged together. While correct, such a ``square-then-average" procedure results in lower signal-to-noise than a hypothetical ``average-then-square" procedure whereby a single power spectrum is formed out of the entire survey. The latter allows the spatial modes of a survey to be averaged together coherently, which allows instrumental noise to be averaged down very quickly. Roughly speaking, if $N$ patches of sky are averaged in a coherent fashion to constrain a particular spatial mode, the noise on the measured mode amplitude averages down as $1/\sqrt{N}$. Squaring this amplitude to form a power spectrum then results in a quicker $1/N$ scaling of noise. In contrast, a ``square-then-average" method combines $N$ independent pieces of information after squaring, and thus the power spectrum noise scales more slowly\footnote{In \citet{parsons_et_al2016} it was shown that in specialized situations it is possible to pre-filter visibility data from an interferometer to recover some of the loss of sensitivity from a square-then-average approach. However, such an approach does not recover large scale angular modes from a wide field of view.} as $1/\sqrt{N}$. The result is a less sensitive statistic, whether for the diagnosis of systematics or for a cosmological measurement. To be fair, one could recover the lost sensitivity by also computing all cross-correlations between a series of small overlapping patches. However, the necessary geometric adjustments for such high precision mosaicking will likely be computationally wasteful, and it quickly becomes preferable to adopt an approach that incorporates the curved sky from the beginning.

In this paper, we rectify the shortcomings of the $k_\perp$-$k_\parallel$ plane by introducing an alternative that is well-defined in the wide-field limit. Rather than expanding sky emission in a basis of rectilinear Fourier modes, we propose a spherical Fourier-Bessel basis. In this basis, the sky brightness temperature $T(\mathbf{r})$ of a survey (where $\mathbf{r}$ is the comoving position) is expressed in terms of $\overline{T}_{\ell m} (k)$, defined as\footnote{
It is an unfortunate coincidence that the spherical harmonic indices are typically denoted by $\ell$ and $m$ in the cosmological literature, while in radio astronomy they are reserved for the direction cosines from zenith in the east-west and north-south directions, respectively. In this paper, $\ell$ and $m$ will always represent spherical harmonic indices, and never direction cosines.}
\begin{equation}
\label{eq:TellmEverything}
\overline{T}_{\ell m} (k) \equiv \sqrt{\frac{2}{\pi}} \int \! d\Omega dr\, r^2 j_\ell (kr) Y_{\ell m}^* (\rhat) T(\mathbf{r}),
\end{equation}
where $k$ is the \emph{total} wavenumber, $\ell$ and $m$ are the spherical harmonic indices, $Y_{\ell m}$ denotes the corresponding spherical harmonic, $r \equiv | \mathbf{r}|$ is the radial distance, $\mathbf{\hat{r}} \equiv \mathbf{r} / r$ is the angular direction unit vector\footnote{In this paper, we use hats for two different purposes. When placed above a vector (e.g., with $\rhat$), the hat indicates that the vector is a unit vector. When placed above a scalar (e.g., with $\widehat{P}$), the hat indicates an estimator of the hatless quantity.}, and $j_\ell$ is the $\ell$th order spherical Bessel function of the first kind. The quantity $P(k_\perp, k_\parallel)$ is replaced by the analogous quantity $S_\ell (k)$, the spherical harmonic power spectrum, which roughly takes the form
\begin{equation}
\label{eq:Sellkrough}
S_\ell (k) \propto \frac{1}{2 \ell + 1} \sum_{m = -\ell}^\ell |\overline{T}_{\ell m} (k)|^2,
\end{equation}
where the sum over $m$ is analogous to the binning of $k_x$ and $k_y$ into $k_\perp$, and a more rigorous definition (with constants of proportionality) will be defined in Section \ref{sec:SphericalPspecFormalism}. Instead of the $k_\perp$-$k_\parallel$ plane, power spectrum measurements are now expressed on an $\ell$-$k$ plane. Now, we will show in Section \ref{sec:SphericalPspecFormalism} that in the limit of a translationally invariant cosmological field, $S_\ell (k)$ reduces to $P(k)$. Therefore, just as $P(k_\perp, k_\parallel)$ can be averaged along contours of constant $k$ to form $P(k)$ once systematic effects are under control, the same can be done for $S_\ell (k)$ to form $P(k)$ by averaging over all values of $\ell$ for a particular $k$.

\begin{figure}[!]
	\centering
	\includegraphics[width=0.35\textwidth,trim=2.0cm 1.2cm 0.8cm 0.5cm,clip]{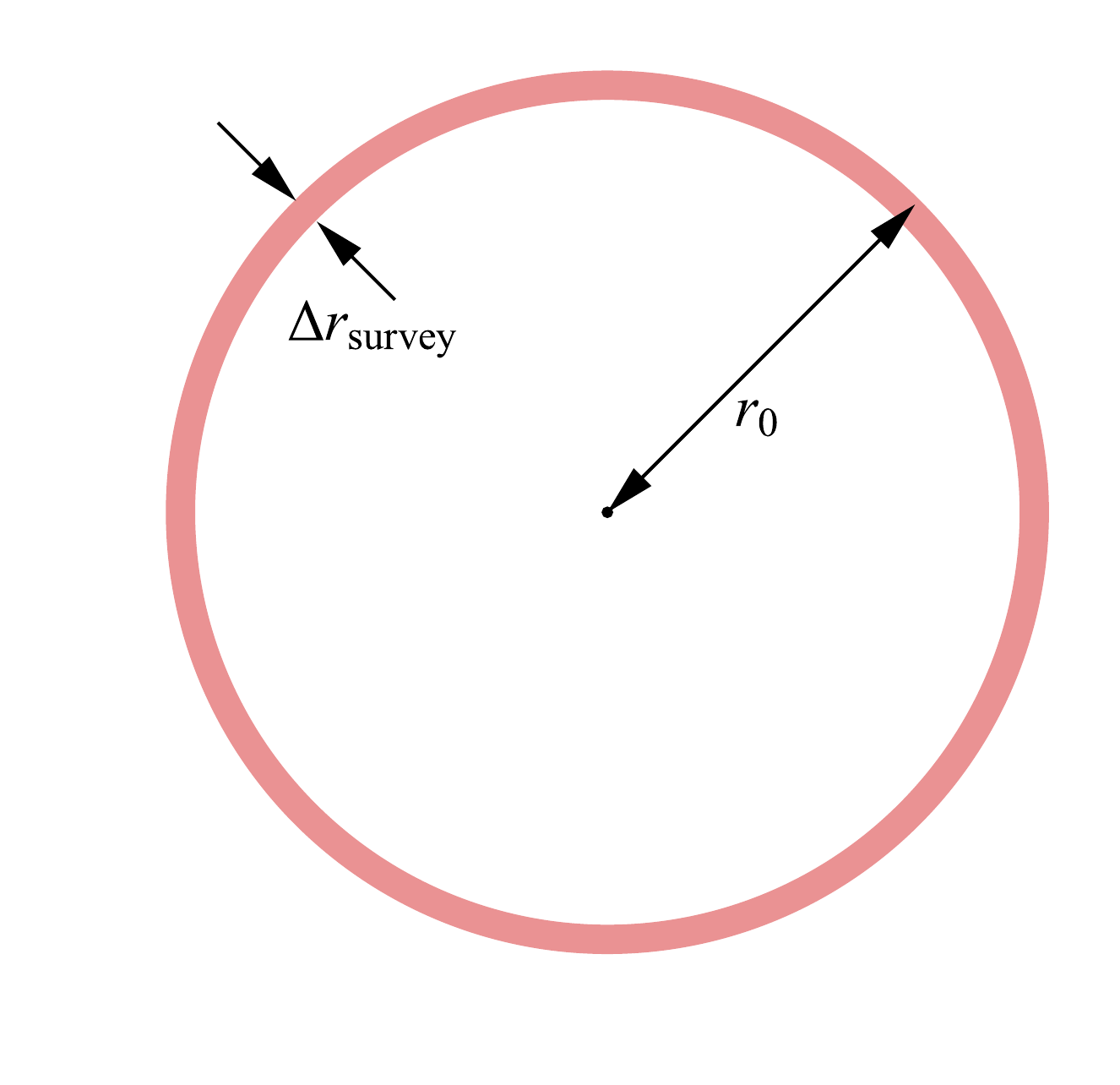}
	\caption{A typical geometry for an intensity mapping survey. The observer is located in the middle, and the thin pink ring denotes the survey's coverage. Particularly at high redshifts (such as those relevant to Epoch of Reionization $21\,\textrm{cm}$ measurements), the cosmological scalings of Section \ref{sec:Notation} typically result in surveys where the median comoving radial distance $r_0$ is much larger than the total radial extent $\Delta r_\textrm{survey}$, i.e., thin-shell surveys. Additionally, most intensity mapping surveys have much higher spectral/radial resolution than they do angular resolution.}
	\label{fig:surveyGeom}
\end{figure}

Spherical Fourier-Bessel methods have been explored in the past within the galaxy survey literature \citep{binney_quinn1991,lahav_et_al1994,fisher_et_al1994,fisher_et_al1995,heavens_taylor1995,zaroubi_et_al1995,castro_et_al2005,leistedt_et_al2012,rassat_refregier2012,shapiro_et_al2012,pratten_munshi2013,yoo_desjacques2013}. In this paper, we build upon these methods and present a framework for implementing them in an analysis of intensity mapping data. We emphasize the way in which intensity mapping surveys have unique geometric properties, and how these properties affect spherical Fourier-Bessel methods. For instance, we pay special attention to the fact that particularly for the highest redshift observations, intensity mapping experiments probe survey volumes that are radially compressed but angularly expansive (as illustrated in Figure \ref{fig:surveyGeom}). In harmonic space, this expectation is reversed, and there is excellent spatial resolution along the line-of-sight (since high spectral resolution is relatively easy to achieve), but poor angular resolution. In addition to addressing these geometric peculiarities, we also show how interferometric data can be analyzed with spherical Fourier-Bessel methods. Importantly, we find that the foregrounds again appear as a wedge in interferometric measurements of $S_\ell (k)$, which suggests that the $\ell$-$k$ plane is at least as powerful a diagnostic tool as the $k_\perp$-$k_\parallel$ plane, particularly given the signal-to-noise advantages discussed above.\footnote{This does not, of course, preclude the examination of systematics in other spaces. For example, though cable reflections may have well-defined signatures on the $k_\perp$-$k_\parallel$ or $\ell$-$k$ planes, they are an example of a systematic that can (and should) also be diagnosed in spaces appropriate for raw data coming off an instrument.}

The rest of this paper is organized as follows. In Section \ref{sec:Notation} we establish notational conventions for this paper. Section \ref{sec:SphericalPspecFormalism} introduces spherical Fourier-Bessel methods for power spectrum estimation, with the complication of finite surveys (in both the angular and spectral directions) the subject of Section \ref{sec:FiniteVolume}. In Section \ref{sec:Foregrounds} we compute the signature of smooth spectrum foregrounds on the $\ell$-$k$ plane. Interloper lines are explored in Section \ref{sec:Interlopers}. A framework for interferometric power spectrum estimation using spherical Fourier-Bessel methods (which includes a derivation of the foreground wedge) is presented in Section \ref{sec:Interferometry}. To build intuition, we develop a parallel series of flat-sky, narrow field-of-view expressions in a series of Appendices. Our conclusions are summarized in Section \ref{sec:Conclusions}. Because of the large number of mathematical quantities defined in this paper, we provide a glossary of important symbols for the reader's convenience in Table \ref{tab:Definitions}.

\begin{table*}
\caption{\label{tab:Definitions}Glossary of important mathematical quantities.  The ``context" column gives equation references, typically either their defining equation or their first appearance in the text.}
\begin{ruledtabular}
\begin{tabular}{lll}
Quantity & Meaning/Definition &  Context \\
\hline
$\mathbf{r}$ & Comoving position & Section \ref{sec:Intro} \\
$\mathbf{\hat{r}}$ & Angular direction unit vector & Section \ref{sec:Intro} \\
$\mathbf{r}_\perp$ & Comoving transverse distance & Eq. \eqref{eq:AngularConversion} \\
$r(\nu)$ or $r_\nu$ & Comoving radial distance & Eq. \eqref{eq:ComovingDistDef} \\
$s(r)$ & Incorrect radial distance assumed for true radial distance $r$ due to interloper lines & Eq. \eqref{sec:Interlopers} \\
$\nu(r)$ or $\nu_r$ & Observed frequency of radio emission & Section \ref{sec:Notation} \\
$\alpha$ & Linearized conversion factor between frequency and radial comoving distance & Eq. \eqref{eq:AlphaConversion} \\
$ \boldsymbol \theta$ & Sky image angle & Eq. \eqref{eq:AngularConversion} \\
$\mathbf{k}$ & Wavevector of rectilinear spatial Fourier modes & Section \ref{sec:Intro} \\
$k_\perp$ & Magnitude of wavevector components perpendicular to line of sight & Section \ref{sec:Intro} \\
$k_\parallel$ & Magnitude of wavevector components parallel to line of sight & Section \ref{sec:Intro} \\
$k$ & Total wavenumber/wavevector magnitude of rectilinear spatial Fourier modes & Section \ref{sec:Intro} \\
$\phi(\mathbf{r})$& Survey volume selection function & Section \ref{sec:FiniteVolume} \\
$\phi(r)$ & Radial survey profile or survey volume selection function assuming full-sky covarage & Section \ref{sec:FiniteVolume} \\
$\Phi(r)$ & Radial survey profile centered on radial midpoint of survey & Section \ref{sec:MostlyRadialNoInterferometry} \\
$T(\mathbf{r})$ or $T(\rhat, \nu)$ & Sky temperature in configuration space & Eq. \eqref{eq:TellmEverything} \\
$ \overline{T}_{\ell m} (k)$ & Sky temperature in spherical Fourier-Bessel space & Eq. \eqref{eq:TellmEverything} \\
$ \overline{T}_{\ell m}^\textrm{meas} (k)$ & Estimated sky temperature in spherical Fourier-Bessel space for finite-volume surveys & Eq. \eqref{eq:Tellm^meas} \\
$\widetilde{T} (\mathbf{k})$ & Sky temperature in rectilinear Fourier space & Eq. \eqref{eq:RectilinearPspecDef} \\
$\kappa (\nu)$ & Frequency spectrum of foreground contaminants & Eq. \eqref{eq:qellk} \\
$q_\ell (k)$ & Frequency spectrum of foreground contaminants in radial spherical Bessel basis & Eq. \eqref{eq:qellk}\\
$a_{\ell m} (\nu)$ & Sky temperature in frequency/spherical harmonic space & Eq. \eqref{eq:SHTdef} \\
$Y_{\ell m} $ & Spherical harmonic function & Section \ref{sec:SphericalPspecFormalism} \\
$\psi_{\ell m} (k; \rhat, \nu)$ & Spherical Fourier-Bessel basis function in configuration space & Eq. \eqref{eq:BasisFcts} \\
$j_\ell (kr) $ & $\ell$th order spherical Bessel function of the first kind & Section \ref{sec:SphericalPspecFormalism} \\
$C_\ell$ & Angular power spectrum & Section \ref{sec:RotationalInvarianceOnly} \\
$P(\mathbf{k})$ & Brightness temperature power spectrum & Section \ref{sec:Intro} \\
$P(k_\perp, k_\parallel)$ & Brightness temperature power spectrum, assuming cylindrical symmetry & Section \ref{sec:Intro} \\
$P(k)$ & Brightness temperature power spectrum, assuming isotropy & Eq. \eqref{eq:RectilinearPspecDef} \\
$S_\ell (k) $ & Spherical harmonic power spectrum & Eq. \eqref{eq:SlkDef} \\
$ \mathbf{b}$ & Interferometer baseline vector & Section \ref{sec:Interferometry} \\
$\tau$ & Interferometric time delay & Eq. \eqref{eq:DelayDef} \\
$V(\mathbf{b}, \nu)$ & Interferometric visibility & Eq. \eqref{eq:VisDef} \\
$\widetilde{V}(\mathbf{b}, \tau)$ & Interferometric visibility in delay space & Eq. \eqref{eq:DelayDef} \\
$A(\rhat, \nu)$ & Primary beam of receiving elements of interferometer & Eq. \eqref{eq:VisDef}\\
$B(\rhat, \nu)$ & Rescaled primary beam & Eq. \eqref{eq:Bdef} \\
$\overline{B^2}(\theta) $ & Squared primary beam profile, averaged azimuthally about a baseline vector & Eq. \eqref{eq:FinalWedgeEq} \\
$\gamma (\nu)$ & Delay transform tapering function & Eq. \eqref{eq:DelayDef} \\
$f_{\ell m} (\mathbf{b}, \nu)$ & Response of baseline $\mathbf{b}$ at frequency $\nu$ to unit perturbation of spherical harmonic mode $Y_{\ell m}$ & Eq. \eqref{eq:flm} \\
$g_{\ell m} (\mathbf{b}, \tau)$ & Response of baseline $\mathbf{b}$ at delay $\tau$ to unit perturbation of spherical harmonic mode $Y_{\ell m}$ & Eq. \eqref{eq:glm} \\
$W_\ell (k; \mathbf{b}, \tau)$ & Spherical harmonic power spectrum window function for a single baseline delay-based & Eq. \eqref{eq:DelayWindowFcts} \\
&  power spectrum estimate & \\
$\Theta(\nu)$  & Re-centered frequency profile of the foregrounds as seen in the data, with finite bandwidth & Section \ref{sec:CurvedSkyWedge} \\
& and tapering effects & \\
$D(\mathbf{r})$ & Survey volume selection function including primary beam, bandwidth, and data analysis & Appendix \ref{sec:RectilinearInterferometerPspecNorm} \\
& tapering effects& \\

\end{tabular}
\end{ruledtabular}
\end{table*}

%
%

\section{Notational preliminaries}
\label{sec:Notation}

Suppose an intensity mapping survey has surveyed the brightness temperature field $T(\rhat, \nu)$ of a particular spectral line as a function of angle (specified here in terms of unit vector $\rhat$) and frequency $\nu$. Such a quantity represents a three-dimensional survey of our Universe, since different frequencies of a spectral line map to different redshifts, and thus different radial distances from the observer. Explicitly, the comoving radial distance $r$ is given by
\begin{equation}
\label{eq:ComovingDistDef}
r (\nu) = \frac{c}{H_0} \int_0^{z(\nu)} \frac{dz^\prime}{E(z^\prime)},
\end{equation}
where $c$ is the speed of light, $H_0$ is the present day Hubble parameter, with
\begin{equation}
1 + z \equiv \frac{\nu_\textrm{rest}}{\nu}\quad \textrm{and} \quad E(z) \equiv \sqrt{\Omega_\Lambda + \Omega_m (1+z)^3},
\end{equation}
where $\nu_\textrm{rest}$ is the rest frequency of the spectral line, $z$ is the redshift, $\Omega_\Lambda$ is the normalized dark energy density, and $\Omega_m$ is the normalized matter density. There is thus a one-to-one mapping between frequency and comoving radial distance, and as shorthand throughout this paper, we will adopt the notation $r_\nu \equiv r(\nu)$. Similarly, we will often use the symbol $\nu_r$ to denote frequency, with the subscript reminding us that the observed frequency is a function of the radial distance. Given a radial distance, transverse distances may also be computed given $\rhat$ (or angle on the sky) using simple geometry.

If one's survey occurs over a narrow radial range, the distance-frequency relation is often replaced by a linearized approximation where
\begin{equation}
\label{eq:LinearDistanceApprox}
r - r_\textrm{ref} \approx - \alpha (\nu - \nu_\textrm{ref} ),
\end{equation}
with $r_\textrm{ref}$ and $\nu_\textrm{ref}$ being a reference comoving radial distance and a reference frequency, respectively, with values constrained by Eq. \eqref{eq:ComovingDistDef}, and
\begin{equation}
\label{eq:AlphaConversion}
\alpha \equiv \frac{1}{\nu_\textrm{rest}} \frac{c}{H_0} \frac{(1+z_\textrm{ref})^2}{E(z_\textrm{ref})},
\end{equation}
where $1 + z_\textrm{ref} = \nu_\textrm{rest} / \nu_\textrm{ref}$. In this paper, the symbols $\nu_r$ and $r_\nu$ will always refer to the exact nonlinear relations, and any invocations of the linearized approximations will be written out explicitly using Eq. \eqref{eq:LinearDistanceApprox}. When using the linearized approximation for the radial distance, we will often (though not always) also make the small angle approximation for converting between the angle $\boldsymbol \theta$ and the transverse comoving position $\mathbf{r}_\perp$ from some reference direction, where
\begin{equation}
\label{eq:AngularConversion}
\mathbf{r}_\perp = r \boldsymbol \theta.
\end{equation}

Given the well-defined prescriptions for converting between instrument-centric parameters (such as frequency $\nu$ and direction on the sky $\rhat$) and cosmology-centric ones (such as $r$ and $\mathbf{r}_\perp$), we will often use both sets of parameters to describe the same quantities. For example, we will sometimes write the brightness temperature field as $T(\rhat, \nu)$, whereas other times we will write the same quantity as $T(\mathbf{r})$, where $\mathbf{r}$ is the comoving position. We will additionally find it useful to exhibit similar flexibility in our notation even for quantities that are not cosmological in nature, such as the primary beam of a radio telescope.

\section{Spherical Fourier-Bessel Formalism}
\label{sec:SphericalPspecFormalism}

In this section we introduce the mathematical framework for describing the sky in terms of the spherical harmonic power spectrum. Our treatment here is essentially identical to that of \citet{yoo_desjacques2013}, albeit with different Fourier-Bessel transform conventions. No claims of originality are made in this section (except perhaps for Section \ref{sec:RotationalInvarianceOnly}), and the formalism is included only for completeness. We will, however, occasionally provide previews of how various parts of the framework are particularly helpful for intensity mapping and interferometry.

In the spherical Fourier-Bessel basis, angular fluctuations are expressed by expanding the temperature field $T(\rhat, \nu)$ in spherical harmonics, such that
\begin{equation}
\label{eq:SHTdef}
a_{\ell m} (\nu) \equiv \int d\Omega Y_{\ell m}^* (\rhat) T(\rhat, \nu).
\end{equation}
To capture modes along the line-of-sight, we perform a Fourier-Bessel transform along the frequency direction, yielding
\begin{equation}
\label{eq:FBdef}
\overline{T}_{\ell m} (k) \equiv \sqrt{\frac{2}{\pi}} \int_0^\infty \! dr\, r^2 j_\ell (kr) a_{\ell m} (\nu_r),
\end{equation}
with these last two expressions of course combining to give Eq. \eqref{eq:TellmEverything}. The temperature field of the sky may therefore be thought of as being a linear combination of a set of basis functions $\psi_{\ell m } (k; \rhat, \nu)$ that are indexed by $(k,\ell,m)$, so that
\begin{equation}
\label{eq:InverseTrans}
T(\rhat, \nu) = \sum_{\ell m} \int dk\, \psi_{\ell m } (k; \rhat, \nu) \overline{T}_{\ell m} (k),
\end{equation}
where
\begin{equation}
\label{eq:BasisFcts}
\psi_{\ell m } (k; \rhat, \nu) \equiv k^2 \sqrt{\frac{2}{\pi}} j_\ell (kr_\nu) Y_{\ell m} (\rhat).
\end{equation}
Eqs. \eqref{eq:SHTdef} and \eqref{eq:FBdef} are the forward transforms into the harmonic basis, while Eqs. \eqref{eq:InverseTrans} and \eqref{eq:BasisFcts} define the inverse transforms back into configuration space. This can be verified by substituting Eq. \eqref{eq:FBdef} into Eq. \eqref{eq:InverseTrans}, and using orthonormality of spherical harmonics, as well as the analogous identity for spherical Bessel functions, given by
\begin{equation}
\label{eq:BesselOrthog}
\int \! dr \,r^2 j_\ell (k r) j_\ell (k^\prime r) = \frac{\pi}{2 k k^\prime} \delta^D (k - k^\prime),
\end{equation}
where $\delta^D$ is the Dirac delta function. Note that our convention for the radial transform differs from that of most works in the literature. From Eqs. \eqref{eq:FBdef} and \eqref{eq:BasisFcts}, one sees that our convention is symmetric in the following sense. Whether one is switching from $r$-space to $k$-space or vice versa, the prescription is always to multiply by $\sqrt{2 / \pi} j_\ell (kr)$ and the square of the coordinate (i.e., $r^2$ or $k^2$) of the original space before integrating over it. This makes our forward and backward transforms aesthetically and conveniently symmetric. Most previous works (e.g., \citealt{leistedt_et_al2012,rassat_refregier2012,yoo_desjacques2013}), in contrast, opt for an asymmetric convention: an extra factor of $k$ is included in the forward transform from $r$ to $k$, and correspondingly there is one fewer factor of $k$ in the backwards transform.

\subsection{Translationally invariant fields in the spherical Fourier-Bessel formalism}
\label{eq:TransInvarFields}
In some sense, the decision to expand fluctuation modes along the line of sight in terms of spherical Bessel functions rather than some other set of basis functions is arbitrary. However, we will now show that spherical Bessel functions are a particularly good choice for describing temperature fields that are statistically translation invariant. Translation-invariant fields admit a representation in terms of their power spectrum $P(k)$, which we define implicitly via the equation\footnote{We implicitly assume throughout this paper that we are dealing only with temperature \emph{fluctuations}. In other words, we assume that that the mean sky temperature has already been subtracted off (or simply does not enter the measurement itself, as is the case with most interferometric measurements).}
\begin{equation}
\label{eq:RectilinearPspecDef}
\langle \widetilde{T} (\mathbf{k}) \widetilde{T}^* (\mathbf{k^\prime}) \rangle = (2 \pi)^3 \delta^D (\mathbf{k} - \mathbf{k}^\prime) P(k),
\end{equation}
where the angled brackets $\langle \cdots \rangle$ signify an ensemble average over random realizations of the cosmological temperature field $T(\mathbf{r})$, whose Fourier transform $\widetilde{T} (\mathbf{k})$ we define by the convention
\begin{equation}
\label{eq:forwardNormal}
\widetilde{T} (\mathbf{k}) = \int \! d^3 r \,e^{-i \mathbf{k} \cdot \mathbf{r}} T(\mathbf{r})
\end{equation}
with the inverse transform given by
\begin{equation}
\label{eq:inverseNormal}
T(\mathbf{r}) = \int \! \frac{d^3 k}{(2 \pi)^3} e^{i \mathbf{k} \cdot \mathbf{r}} \widetilde{T} (\mathbf{k}).
\end{equation}
Unless otherwise stated, this Fourier convention for the temperature field will be the one used for all Fourier transforms in this paper. Ideally, our spherical Fourier-Bessel description should be directly relatable to $P(k)$, for it would be pointless if an estimation of the power spectrum required first returning to position space. We will now show that this requirement is met by our $\overline{T}_{\ell m} (k)$ modes.

To relate $\overline{T}_{\ell m} (k)$ to $P(k)$, we combine Eqs. \eqref{eq:SHTdef}, \eqref{eq:FBdef}, and \eqref{eq:inverseNormal} to obtain
\begin{equation}
\label{eq:YetAnotherTellm}
\overline{T}_{\ell m} (k) = \sqrt{\frac{2}{\pi}} \int \! \frac{d^3 k^\prime}{(2 \pi)^3} \widetilde{T} (\mathbf{k}^\prime) \int \! d^3 r\, j_\ell (kr) Y_{\ell m}^*(\rhat) e^{i \mathbf{k}^\prime \cdot \mathbf{r}}.
\end{equation}
To simplify this, we expand $e^{i \mathbf{k}^\prime \cdot \mathbf{r}}$ in spherical harmonics using the identity
\begin{equation}
\label{eq:PlaneWaveSphericalHarmonicExpansion}
e^{i \mathbf{k} \cdot \mathbf{r}} = 4\pi \sum_{\ell m} i^\ell j_\ell (kr) Y_{\ell m}^* (\khat) Y_{\ell m} (\rhat),
\end{equation}
which leads to
\begin{equation}
\label{eq:TlmTkConversion}
\overline{T}_{\ell m} (k) = \frac{i^\ell}{(2\pi)^{\frac{3}{2}}} \int \frac{d^3 k^\prime}{k k^\prime} Y_{\ell m}^* (\khat^\prime) \delta^D (k - k^\prime) \widetilde{T} (\mathbf{k}^\prime).
\end{equation}
This provides a link between the temperature field as expressed in our $(k,\ell, m)$ basis, and the same field in the rectilinear Fourier basis. Taking a cue from Eq. \eqref{eq:RectilinearPspecDef}, where the power spectrum is closely related to the two-point correlation between different rectilinear Fourier modes, we may form a two-point correlator between different modes in our spherical Fourier-Bessel basis, giving
\begin{eqnarray}
\label{eq:CurvedPspecDef}
\langle \overline{T}_{\ell m} (k) \overline{T}_{\ell^\prime m^\prime}^* (k^\prime) \rangle && = \frac{i^\ell (-i)^{\ell^\prime}}{(2\pi)^3} \!\! \int \frac{d^3 k_1}{k k_1} \frac{d^3 k_2}{k^\prime k_2}  \nonumber \\
&& \qquad \times Y_{\ell m}^* (\khat_1) Y_{\ell^\prime m^\prime} (\khat_2) \langle \widetilde{T} (\mathbf{k}_1) \widetilde{T} (\mathbf{k}_2)^* \rangle\nonumber \\
&& \qquad \times \delta^D (k - k_1) \delta^D (k^\prime - k_2)   \nonumber \\
&& = \frac{\delta^D(k - k^\prime) }{k^2} \delta_{\ell \ell^\prime} \delta_{m m^\prime} P(k),
\end{eqnarray}
where the last equality follows from Eq. \eqref{eq:RectilinearPspecDef} and some algebraic simplifications. From this, we see that forming the power spectrum from $\overline{T}_{\ell m} (k)$ modes is remarkably similar to forming it from the rectilinear Fourier modes. Comparing Eqs. \eqref{eq:RectilinearPspecDef} and \eqref{eq:CurvedPspecDef}, we see that if (roughly speaking) one can form $P(k)$ by squaring $\widetilde{T} (\mathbf{k})$ and normalizing appropriately, one can equally well form $P(k)$ by squaring $\overline{T}_{\ell m} (k)$ and normalizing (albeit with a different---and $k$ dependent---normalization that we will derive more explicitly in Section \ref{sec:FiniteVolume}).

To understand why the squaring of $\overline{T}_{\ell m} (k)$ produces such a similar result to squaring $\widetilde{T} (k)$ (with both giving a result proportional to the power spectrum), notice that Eq. \eqref{eq:TlmTkConversion} can be simplified to give
\begin{equation}
\overline{T}_{\ell m} (k) = \frac{i^\ell}{(2\pi)^{\frac{3}{2}}} \int d\Omega_k Y^*_{\ell m} (\khat)  \widetilde{T} (\mathbf{k})\bigg{|}_{|\mathbf{k}| = k},
\end{equation}
where $ \widetilde{T} (\mathbf{k})$ is restricted to the shell where $|\mathbf{k}| = k$. In this form, one sees that an alternate way to understand our spherical harmonic Bessel modes is to view them as a spherical harmonic decomposition of $ \widetilde{T} (\mathbf{k})$ in Fourier space. In other words, going from the rectilinear Fourier modes to spherical harmonic Bessel modes is simply a change of basis---to spherical harmonics---in angular Fourier coordinates. Now, suppose one were to form an estimate of $P(k)$ in by squaring $ \widetilde{T} (\mathbf{k})$ and then averaging over a shell of constant $|\mathbf{k}| = k$. Parseval's theorem ensures that such a squaring and averaging operation is basis-independent. Thus, it does not matter whether the Fourier amplitudes on the shell of constant $|\mathbf{k}| = k$ are expressed in a spherical harmonic basis. Squaring and averaging $\overline{T}_{\ell m} (k)$ must therefore also yield the power spectrum, up to some $k$-dependent conversion factors to account for the radius of shells in Fourier space. Note that Eq. \eqref{eq:CurvedPspecDef} also cements the interpretation (suggested by our notation) that the quantity $k$ of our Fourier-Bessel basis is the total magnitude of the wavevector $\mathbf{k}$, rather than some wavenumber that only pertains to radial fluctuations.


\subsection{Rotationally invariant fields in the spherical Fourier-Bessel formalism}
\label{sec:RotationalInvarianceOnly}

While the cosmological temperature field is expected to possess translationally invariant statistics, contaminants in an intensity mapping survey (such as foreground emission) will in general not possess such symmetry. This difference in symmetry will result in different signatures on the $\ell$-$k$ plane that can in principle be used to separate contaminants from the cosmological signal.

To elucidate the contrast in these signatures, suppose we discard the assumption (from previous derivations) of translationally invariant statistics. In general, the two-point correlator will cease to exhibit the diagonal form given by Eq. \eqref{eq:CurvedPspecDef}. As a concrete example of this, consider a random temperature field that is statistically isotropic but not homogeneous. In the radial direction, suppose this field has some fixed (non-random and angular position-independent) radial dependence. Such a field would be an appropriate description for a (hypothetical) population of unresolved point sources. Under these assumptions, Eq. \eqref{eq:FBdef} reduces to
\begin{equation}
\overline{T}_{\ell m} (k) = a_{\ell m} q_\ell (k),
\end{equation}
where
\begin{equation}
\label{eq:qellk}
q_\ell (k) \equiv \sqrt{\frac{2}{\pi}} \int_0^\infty dr r^2 j_\ell (kr) \kappa (\nu_r),
\end{equation}
with $\kappa (\nu_r)$ specifying the spectral (and therefore radial) dependence of our hypothetical sky as it appears in our data. The two-point correlator then becomes
\begin{equation}
\langle \overline{T}_{\ell m} (k) \overline{T}_{\ell^\prime m^\prime}^* (k^\prime) \rangle  = C_\ell q_\ell(k) q_\ell (k^\prime) \delta_{\ell \ell^\prime} \delta_{m m^\prime},
\end{equation}
where statistical rotation invariance of the field allows us to invoke relation $\langle a_{\ell m} a_{\ell^\prime m^\prime}^* \rangle \equiv C_\ell \delta_{\ell \ell^\prime} \delta_{m m^\prime}$, with $C_\ell$ signifying the angular power spectrum.

Our example illustrates the way in which the two-point correlator ceases to be diagonal in $k$ and $k^\prime$ once translation invariance is broken. In general, if the sky exhibits rotational invariance (in the statistical sense), the correlator takes the form
\begin{equation}
\langle \overline{T}_{\ell m} (k) \overline{T}_{\ell^\prime m^\prime}^* (k^\prime) \rangle \equiv M_\ell (k, k^\prime) \delta_{\ell \ell^\prime} \delta_{m m^\prime},
\end{equation}
for some function $M_\ell (k, k^\prime)$. In the limit that the sky is statistically homogeneous in addition to isotropic, $M_\ell (k, k^\prime)$ becomes $\ell$-independent and reduces to $P(k) \delta^D (k - k ^\prime) / k^2$, as demonstrated in Eq. \eqref{eq:CurvedPspecDef}. If one is simply squaring $\overline{T}_{\ell m} (k)$ measurements to estimate the power spectrum but there are non-statistically homogeneous contaminants in the data, one obtains 
\begin{equation}
\langle | \overline{T}_{\ell m} (k)|^2 \rangle \equiv M_\ell (k) \delta_{\ell \ell^\prime} \delta_{m m^\prime},
\end{equation}
where $M_\ell (k)$ is a function of both $\ell$ and $k$ rather than just $k$ alone.

We thus see that the spherical Fourier-Bessel formulation fulfills the goals we laid out near the beginning of this section. In particular, the foreground contaminants appear differently on the $\ell$-$k$ plane than the cosmological signal does, owing to the translation-invariant statistics of the latter. This generalizes the symmetry arguments for foreground mitigation laid out in \citet{morales_and_hewitt2004} in a way that is well-defined for wide fields of view. We note, however, that as the formalism currently stands, $M_\ell (k)$ and $P(k)$ are not directly comparable; indeed, they have different units. This arises because the two quantities scale differently with volume. For a random cosmological field described by $P(k)$, the magnitude of $\overline{T}_{\ell m} (k)$ scales as $\sqrt{V}$, where $V$ is the volume of a survey. On the other hand, contaminants may not be describable as random fields. In the case of foregrounds, for example, the signal is smooth and coherent along the radial/frequency direction. As a result, $\overline{T}_{\ell m} (k)$ scales more quickly than $\sqrt{V}$. Indeed, the difference between these scalings was proposed as a method for distinguishing between foreground contamination and cosmological signal in \citet{cho_et_al2012}. To derive a quantity for describing survey contaminants  on the $\ell$-$k$ that is directly comparable to $P(k)$ it is necessary to specify a survey volume. In the following sections, we will depart from the idealized treatment considered in this section, where we imagined having access to a perfectly sampled field over an infinite volume.

\section{Estimating the power spectrum from finite-volume surveys}
\label{sec:FiniteVolume}

In this section, we consider the effects of the necessarily finite extent of any real survey. Finite selection effects were considered in \citet{rassat_refregier2012} and \citet{leistedt_et_al2012}, and here we provide a complementary treatment that is not only tailored for intensity mapping, but also provides explicit expressions for the power spectrum on the $\ell$-$k$ plane.

Suppose the extent of our survey is given by a function $\phi(\mathbf{r})$, such that $\phi(\mathbf{r})$ is zero everywhere beyond the boundaries of the survey. A survey with uniform sensitivity can then be modeled by setting $\phi(\mathbf{r}) = 1$ inside the survey. In what follows, however, we do not make this assumption, and we allow for spatially varying sensitivity within the survey. This permits the treatment of angular masks as well as radial selection functions. In general, the temperature field that is analyzed is $\phi(\mathbf{r}) T(\mathbf{r})$ rather than $T(\mathbf{r})$. A result, the measured spherical Fourier-Bessel modes $\overline{T}_{\ell m}^\textrm{meas}(k)$ are not described by Eq. \eqref{eq:TlmTkConversion}, but instead are given by
\begin{eqnarray}
\label{eq:Tellm^meas}
\overline{T}_{\ell m}^\textrm{meas} (k) = \frac{i^\ell}{(2\pi)^{\frac{3}{2}}} \int \frac{d^3 k^\prime}{k k^\prime} && \frac{d^3 k^{\prime \prime}}{(2\pi)^3} Y_{\ell m}^* (\khat^\prime) \delta^D (k - k^\prime) \nonumber \\
&& \times \widetilde{\phi} (\mathbf{k}^\prime - \mathbf{k}^{\prime \prime}) \widetilde{T} (\mathbf{k}^{\prime\prime}),
\end{eqnarray}
where we have invoked the convolution theorem to write our expression in terms of $\widetilde{\phi}$, the Fourier transform of $\phi$.

Despite this revised expression, one might still expect the power spectrum to be closely related to $\overline{T}_{\ell m}^\textrm{meas} (k)$. Squaring and taking the ensemble average gives
\begin{eqnarray}
\langle | \overline{T}_{\ell m}^\textrm{meas} (k) |^2 \rangle 
= \frac{1}{(2\pi)^3} \int \frac{d^3 k_a}{k k_a} \frac{d^3 k_b}{k k_b} \frac{d^3 k_c}{(2\pi)^3} Y_{\ell m}^* (\khat_a) Y_{\ell m} (\khat_b)  \nonumber \\
\times P(k_c) \widetilde{\phi} (\mathbf{k}_a - \mathbf{k}_c) \widetilde{\phi}^* (\mathbf{k}_b - \mathbf{k}_c) \delta^D (k - k_a) \delta^D (k - k_b),\qquad
\end{eqnarray}
where we have again used the definition of the power spectrum from Eq. \eqref{eq:RectilinearPspecDef} to simplify the ensemble average of the two factors of $\widetilde{T}$. Now, if the survey volume is reasonably large, $\phi(\mathbf{r})$ will tend to be a relatively broad function, and thus the two copies of $\widetilde{\phi}$ will be sharply peaked about $\mathbf{k}_a \approx \mathbf{k}_b \approx \mathbf{k}_c$. These then work in conjunction with the two Dirac delta functions to require $k \approx k_c$. With all these conditions, the only part of the integrand that contributes substantially to the integral is the part where $P(k_c) \approx P(k)$, allowing the power spectrum to be factored out of the integral (assuming it is a reasonably smooth function). Doing so and subsequently re-expressing $\widetilde{\phi}$ in terms of $\phi$, our expression simplifies to
\begin{eqnarray}
\label{eq:TlmPkProportionality}
\langle | \overline{T}_{\ell m}^\textrm{meas} (k) |^2 \rangle && \approx  \frac{P(k)}{(2\pi)^3} \int d^3 r \phi^2 (\mathbf{r}) \nonumber \\
&&  \quad \times \Bigg{|} \int \frac{d^3 k_a}{k k_a}Y_{\ell m}^* (\khat_a) e^{-i \mathbf{k}_a \cdot \mathbf{r}} \delta^D (k - k_a) \Bigg{|}^2 \nonumber \\
&& = P(k) \frac{2}{\pi} \int d^3 r \phi^2 (\mathbf{r}) j_\ell^2 (kr) \big{|} Y_{\ell m} (\rhat) \big{|}^2, \qquad
\end{eqnarray}
where in the last equality we performed the integral over $k_a$ by inserting Eq. \eqref{eq:PlaneWaveSphericalHarmonicExpansion} and invoking the orthonormality of spherical harmonics. The final result is a direct proportionality between the ensemble average of hypothetical noiseless measurements $ | \overline{T}_{\ell m}^\textrm{meas} (k) |^2$ and the power spectrum. Heuristically, this equation implies that the power spectrum can be estimated using any $(k,\ell, m)$ mode simply by taking $| \overline{T}_{\ell m}^\textrm{meas} (k) |^2$ and dividing out by everything on the right hand side\footnote{Note that even though this was derived assuming that $P(k)$ is smooth (which does not necessarily hold when substantial foreground contaminants are involved; \citealt{liu_et_al2014b}), the resulting normalization is still the correct one to use.} after $P(k)$. A subsequent averaging of such estimates obtained from modes with the same $k$ but different $\ell$ and $m$ increases the signal-to-noise.

A similar proportionality exists within the framework of rectilinear Fourier modes for relating the squares of the measured Fourier amplitudes $\widetilde{T}^\textrm{meas} (\mathbf{k})$ and $P(k)$ (which we derive in Appendix \ref{sec:RectilinearFKP} to facilitate the comparative discussion that follows). With rectilinear modes, $\langle |\widetilde{T}^\textrm{meas} (\mathbf{k}) |^2 \rangle$ is also proportional to $P(k)$, with the constant of proportionality also given by an integral that has units of volume. However, there exists a crucial difference between the volume integral seen here and the one for the rectilinear framework in Appendix \ref{sec:RectilinearFKP}. With the rectilinear case, the volume factor is independent of the orientation of $\mathbf{k}$ (i.e., $\khat$), so that Fourier modes of all orientations are equally sensitive to the power spectrum. It follows that an optimal estimate of the power spectrum can be obtained by an average of $|\widetilde{T}^\textrm{meas} (\mathbf{k}) |^2$ over spheres of constant $| \mathbf{k}| = k$ with uniform weighting, as we show in Appendix \ref{sec:RectilinearFKP}.

In contrast, the volume integral in Eq. \eqref{eq:TlmPkProportionality} is a function of $\ell$ and $m$. For a particular $(k, \ell, m)$ mode, the value of $\ell$ determines how much the total wavenumber $k$ is comprised of angular fluctuations (as opposed to radial fluctuations), while the value of $m$ determines the orientation of the angular fluctuations. Putting these facts together, it follows that with $\overline{T}_{\ell m}^\textrm{meas} (k)$ modes, the sensitivity to the power spectrum does depend strongly to a mode's orientation. As an example, suppose the survey's sensitivity $\phi(\mathbf{r})$ is localized in small region around some radius $r_0$ away from the observer (illustrated in Figure \ref{fig:surveyGeom}), as is typical for many high-redshift intensity mapping surveys. Now consider (as an extreme case), modes where $\ell \gg k r_0$. For such modes, the Bessel function in Eq. \eqref{eq:TlmPkProportionality} can be approximated by a power series as 
\begin{equation}
j_\ell (kr) \approx \frac{(kr)^\ell}{(2 \ell + 1)!!}.
\end{equation}
The integral on the right hand side of Eq. \eqref{eq:TlmPkProportionality} thus becomes extremely suppressed by a $[(2 \ell + 1)!!]^2$ dependence, giving a small proportionality constant between $| \overline{T}_{\ell m}^\textrm{meas} (k) |^2$ and $P(k)$ for high $\ell$. Thus, high $\ell$ modes that satisfy $\ell \gg k r_0$ are not high signal-to-noise probes of the power spectrum. To understand this, consider instead the modes with $k \sim \ell / r_0$. Such modes are essentially constant in the radial direction, and describe fluctuations that are almost entirely in the angular direction. Temporarily invoking the language of the flat-sky approximation for the sake of intuition, we may say that in this regime, the total wavenumber $k$ is dominated by $k_\perp$. Increasing $\ell$ beyond this to get back to the case where $\ell \gg k r_0$, we have situation that approximately corresponds to having $k_\perp > k$. Such a scenario would be a mathematical impossibility in the flat-sky approximation, and formally the amplitude of the signal would go to zero. In our curved-sky treatment, however, we see that the cut-off for high $\ell$, while dramatic, is not precisely zero. This is due to projection effects, which cause any given $\ell$ mode to sample a spread of $k$ modes, in principle allowing arbitrarily high $\ell$ modes to have some (tiny) response to Fourier modes with very low $k$ values.

With such a strong dependence in power spectrum sensitivity to the values of $\ell$ and $m$, different modes should be weighted differently when averaged together. In principle, this weighting should depend on both $\ell$ and $m$. For simplicity, we will assume that different $m$ values are averaged together with uniform weights. This is a reasonable approximation for wide-field surveys, which is of course the regime that is being targeted in this paper. Indeed, for an all-sky survey, one can show that the integral in Eq. \eqref{eq:TlmPkProportionality} becomes independent of $m$, implying equal sensitivity to all $m$ modes and thus no reason to favor one specific mode over another. Performing the uniform average over Eq. \eqref{eq:TlmPkProportionality} and invoking Uns\"{o}ld's theorem then gives
\begin{equation}
\label{eq:TotallyUnsold}
\frac{\sum_{m = -\ell}^\ell\langle | \overline{T}_{\ell m}^\textrm{meas} (k) |^2 \rangle}{2\ell + 1} \approx \frac{P(k)}{2 \pi^2} \int d^3 r \phi^2 (\mathbf{r}) j_\ell^2 (kr).
\end{equation}
From this, it follows that given a set of modes with some particular $k$ and $\ell$ values, an estimator of the power spectrum can be formed by computing
\begin{equation}
\label{eq:SlkDef}
S_\ell (k) \equiv 2 \pi^2 \left[\int d^3 r \phi^2 (\mathbf{r}) j_\ell^2 (kr)\right]^{-1} \frac{\sum_{m} | \overline{T}_{\ell m}^\textrm{meas} (k) |^2}{2 \ell + 1},
\end{equation}
which we dub the spherical harmonic power spectrum. This is the quantity that we were seeking in Section \ref{sec:RotationalInvarianceOnly}, a curved sky analog to the cylindrical power spectrum $P(k_\perp, k_\parallel)$. If $\overline{T}_{\ell m}^\textrm{meas} (k)$ consists of contaminants to one's measurement, $S_\ell (k)$ would essentially be the ``power spectrum of contaminants", even though such a quantity is in principle not well-defined as the contaminants are typically not statistically translation-invariant. However, $S_\ell (k)$ and $P(k)$ can be directly compared since the two quantities have the same units, and in the limit of translation invariance, the ensemble average of $S_\ell(k)$ reduces to $P(k)$, by construction. We thus have a well-defined quantity that can be considered ``the power spectrum of the signal on the $\ell$-$k$ plane", regardless of the relative ratios of cosmological signal and contaminants.\footnote{As expected from Section \ref{sec:RotationalInvarianceOnly}, the definition of $S_\ell (k)$ depends on the survey geometry $\phi(\mathbf{r})$. This dependence cancels out for the cosmological signal, but not for contaminants. Thus, while two different surveys should give identical results for the cosmological power spectrum $P(k)$, the contaminant (e.g., foreground) contributions to the power are not directly comparable, and two surveys with identical contaminating influences but different sky coverage may measure different total power spectra. Note that this argument is due purely to the differences in scaling with survey volume discussed in Section \ref{sec:RotationalInvarianceOnly}. It thus applies equally well to both $P(k_\perp, k_\parallel)$ and $S_\ell (k)$, and is not simply a peculiarity of the latter.}

Once $S_\ell (k)$ has been computed for all $\ell$ values accessible to an experiment, different $\ell$ modes can be averaged together form a final estimate $\widehat{P} (k)$ of the power spectrum $P(k)$. Unlike with the average over $m$, uneven weights for the $\ell$ average are crucial since different $\ell$ modes can have very different sensitivities to the power spectrum, as our earlier example illustrated. The optimal weights $w_\ell$ for different $\ell$ values will in general depend on the details of one's survey instrument. As a simple toy example, suppose an instrument has equal noise in all $\overline{T}_{\ell m}(k)$ modes (which is an impossibility in practice, since all instruments have finite angular resolution). An optimal signal-to-noise weighting of $| \overline{T}_{\ell m}^\textrm{meas} (k) |^2$ then reduces to a weighting by the strength of the signal, since the noise is constant. This is given by the integral in Eq. \eqref{eq:TotallyUnsold}, which quantifies the extent to which the power spectrum is amplified (or depressed) in each $| \overline{T}_{\ell m}^\textrm{meas} (k) |^2$ mode. Forming a minimum variance estimator then requires a variance (i.e., squared) weighting by this factor, giving an estimator $\widehat{P} (k)$ of the power spectrum that takes the form
\begin{equation}
\label{eq:WeightedPk}
\widehat{P} (k) \equiv \sum_\ell w_\ell S_\ell (k),
\end{equation}
where
\begin{equation}
\label{eq:MinVarEllWeights}
w_\ell \equiv \frac{ \left[ \int d^3 r \phi^2 (\mathbf{r}) j_{\ell}^2 (kr) \right]^2}{\sum_{\ell^\prime} \left[ \int d^3 r^\prime \phi^2 (\mathbf{r}^\prime) j_{\ell^\prime}^2 (kr^\prime) \right]^2}.
\end{equation}


\section{Foreground signatures in the spherical harmonic power spectrum}
\label{sec:Foregrounds}

Having established $S_\ell (k)$ as a potential tool for separating contaminants from cosmological signal in a power spectrum measurement, we now specialize and consider the particular case of astrophysical foreground contamination. Our goal is to derive the signature of foreground contamination in $S_\ell (k)$, and to show that $S_\ell (k)$ is indeed a useful diagnostic for separating foregrounds from the cosmological signal. We will find that $S_\ell (k)$ performs this role for wide-field, curved-sky power spectrum analyses just as well as $P(k_\perp, k_\parallel)$ did for narrow fields of view. By this, we mean that in both cases the foregrounds are localized to predictable regions in the $\ell$-$k$ or $k_\perp$-$k_\parallel$ plane, enabling foregrounds to be mitigated by a few simple cuts to data.

\begin{figure}[!]
	\centering
	\includegraphics[width=0.48\textwidth]{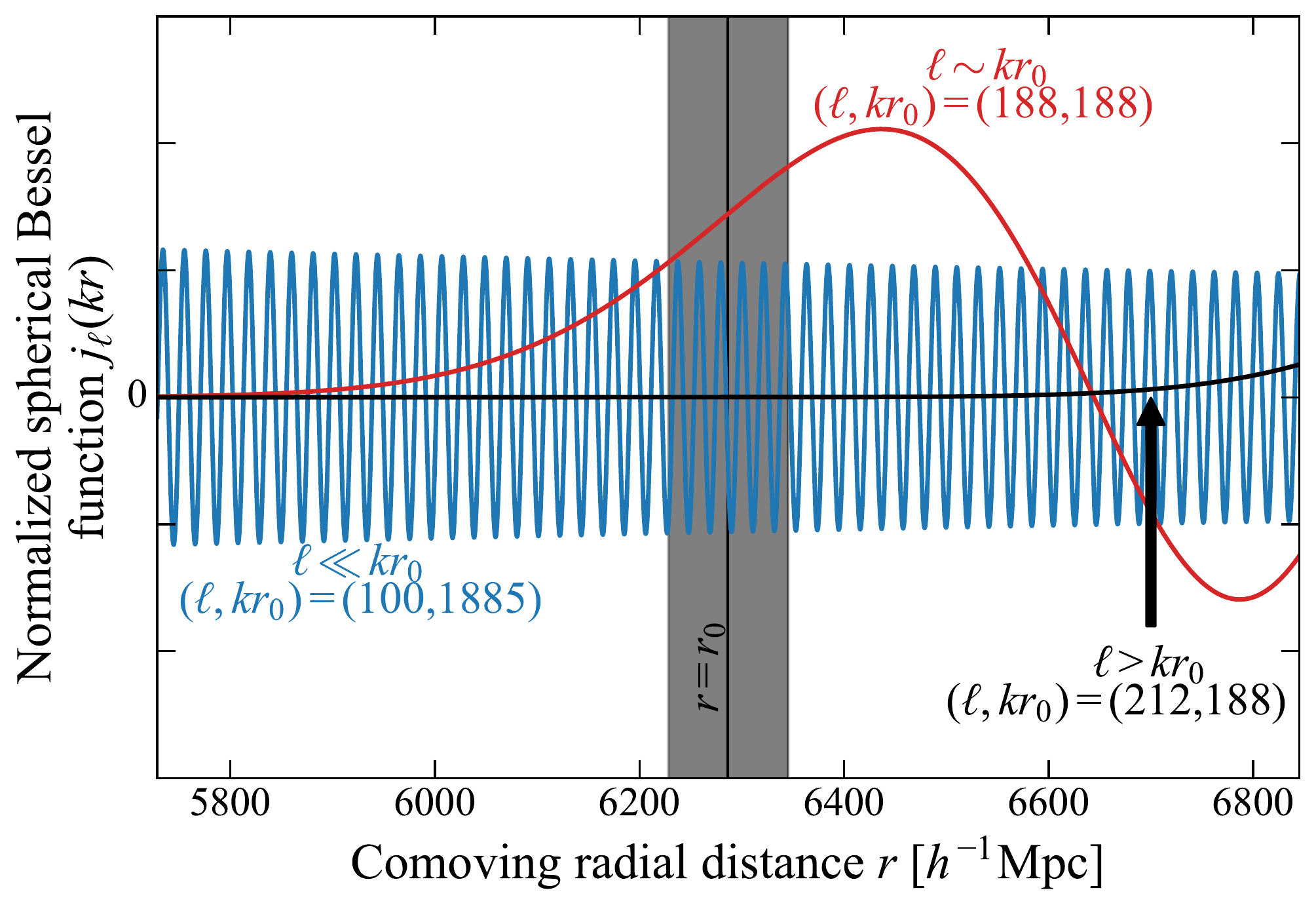}
	\caption{Example spherical Bessel functions $j_\ell (kr)$, arbitrarily normalized for ease of comparison. The grey band indicates the comoving radial extent of a $21\,\textrm{cm}$ intensity mapping survey operating from $145\,\textrm{MHz}$ to $155\,\textrm{MHz}$ (corresponding to a central redshift of 8.5, or a central radial distance of $r_0 \approx 6290h^{-1}$Mpc). The spherical Bessel functions enter in the radial transform from position space to the spherical Fourier-Bessel basis, and are integrated over the grey band with an $r^2$ weighting. Basis functions that describe fluctuations that are predominantly in the angular directions have $\ell \sim kr_0$ behave as power laws over the radial profile of the survey (red curve), and essentially average over the line-of-sight direction. Those whose fluctuations are oriented mainly in the radial direction have $\ell \lesssim kr_0$ behave like slowly modulated sinusoids (blue curve), and effectively take a Fourier transform along the line of sight. Modes with $\ell > kr_0$ (black curve) have very little response.}
	\label{fig:bessels}
\end{figure}

\begin{figure*}[!]
	\centering
	\includegraphics[width=1.0\textwidth]{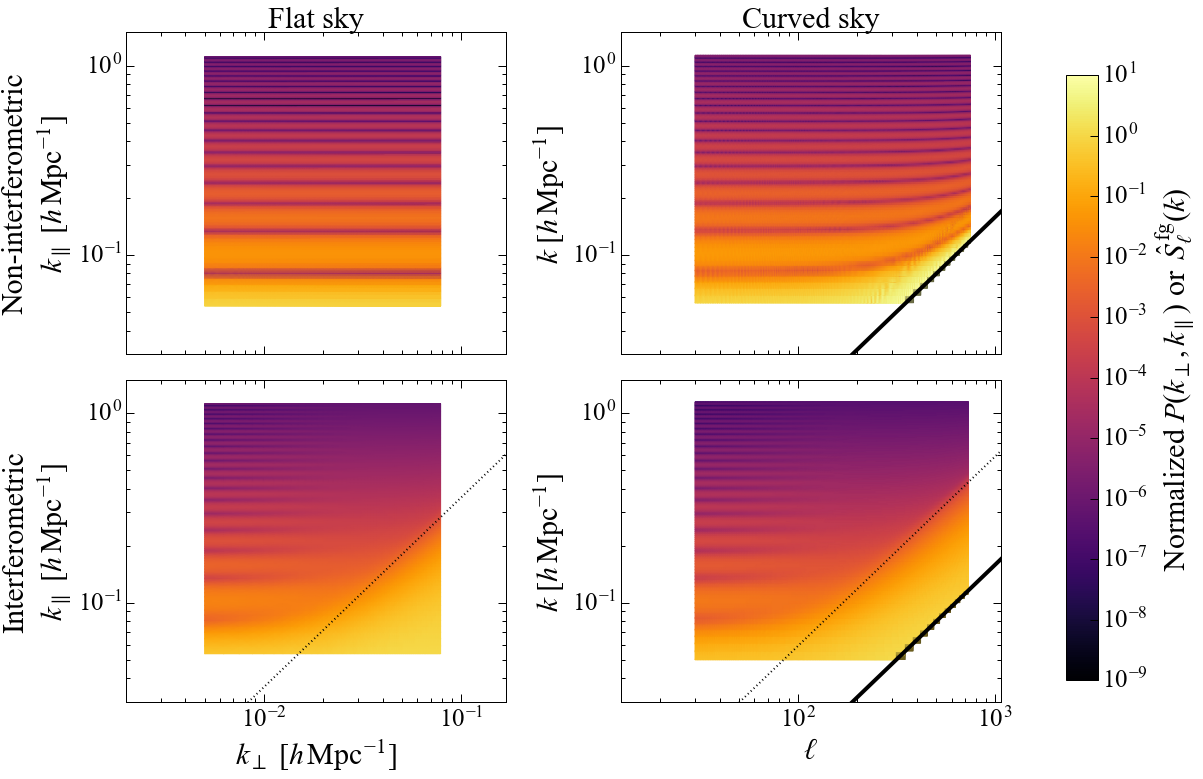}
	\caption{Smooth spectrum foreground signatures, plotted as $\widehat{P}^\textrm{fg}(k_\perp, k_\parallel)$ on the $k_\perp$-$k_\parallel$ (left column) for the flat-sky treatment, and as $S^\textrm{fg}_\ell (k)$ on the $\ell$-$k$ plane (right column). The top row shows the inherent foreground signature (which is roughly what is seen by a non-interferometric instrument), while the bottom row shows the signature seen by an interferometer. One sees that the chromatic response of an interferometer causes foregrounds to leak upwards on the $k_\perp$-$k_\parallel$ and $\ell$-$k$ planes, but that this leakage is confined to a foreground wedge region. Thin dotted lines show the theoretically expected extent of the foreground wedge, the boundary of which is given by Eq. \eqref{eq:WedgeLineEquation} for the flat sky and Eq. \eqref{eq:Finalkellwedge} for the curved sky. The thick black line on the curved-sky plots demarcate the $\ell = k r_0$ boundary, below which there is very little power to be measured. The flat sky plots have been normalized to their values at $(k_\perp, k_\parallel) \sim (0.04\,h\textrm{Mpc}^{-1},0.05\,h\textrm{Mpc}^{-1})$ and the curved sky plots have been normalized to their values at $(\ell,k) \sim (230,0.05\,h\textrm{Mpc}^{-1})$.}
	\label{fig:fgSigs}
\end{figure*}

When performing an intensity mapping survey with a spectral line, the cosmological component of the signal is expected to fluctuate rapidly as a function of frequency, since different frequencies probe different portions of our Universe. Foregrounds, on the other hand, are expected to be spectrally smooth \citep{dimatteo_et_al2002,oh_and_mack2003,deOliveiraCosta_et_al2008,jelic_et_al2008,liu_and_tegmark2012}. In principle, this allows foregrounds to be separated from the cosmological signal, for instance by fitting out a smooth spectral component \citep{wang_et_al2006,liu_et_al2009a,bowman_et_al2009,liu_et_al2009b}. To take an even simpler approach, one expects spectrally smooth foregrounds to appear only at low $k_\parallel$, since $k_\parallel$ is the Fourier dual to line-of-sight distance, which is probed by the frequency spectrum. This is illustrated in the top left panel of Figure \ref{fig:fgSigs}, where we compute the $P(k_\perp, k_\parallel)$ signature of flat spectrum foregrounds for an intensity mapping survey with a radial profile given by
\begin{equation}
\label{eq:CosineRadial}
\phi(r) = \cos \left[ \pi \left( \frac{r-r_0}{r_\textrm{max} - r_\textrm{min}} \right) \right],
\end{equation}
within the comoving radial range of $r_\textrm{min} \approx 6230\,h^{-1}\textrm{Mpc}$ to $r_\textrm{max} \approx 6350\,h^{-1}\textrm{Mpc}$ and zero outside this range. This is representative of a $21\,\textrm{cm}$ intensity mapping survey with a $10\,\textrm{MHz}$ bandwidth centered around a frequency of $150\,\textrm{MHz}$ (corresponding roughly to $z \sim 8.5$). The precise form of the profile is arbitrary, and is only for illustrative purposes in this paper. In the angular direction we assume all-sky coverage. The foregrounds are assumed to have intrinsically flat (frequency-independent) spectra. One sees that their contribution to the power spectrum decreases in amplitude rapidly towards higher $k_\parallel$, suggesting that foregrounds can be mostly avoided by simply looking away from the lowest $k_\parallel$. Note that we have arbitrarily normalized the power to emphasize the morphology (rather than the absolute level) on the $k_\perp$-$k_\parallel$ plane.

We now generalize the signature of foregrounds from the narrow-field to the curved sky using the spherical harmonic power spectrum. The foregrounds are again assumed to be independent of frequency, giving rise to a set of frequency-independent spherical harmonic coefficients $a^\textrm{fg}_{\ell m}$. The resulting $(k,\ell, m)$ modes are then given by
\begin{equation}
\label{eq:fgTlm}
\overline{T}_{\ell m}^\textrm{fg} (k) = a_{\ell m}^\textrm{fg} \sqrt{\frac{2}{\pi}} \int_0^\infty \!dr\, r^2 j_\ell (kr) \phi(r),
\end{equation}
which is simply Eq. \eqref{eq:FBdef} but with the limitation of a survey volume $\phi$ and a flat spectrum assumption. Note that in this section, we will assume that the survey covers the entire angular extent of the sky (as depicted in Figure \ref{fig:surveyGeom}), so that we have $\phi(r) $ rather than $\phi (\mathbf{r})$. In an analysis of real data this assumption may be inappropriate, but here we invoke it for the purposes of mathematical clarity. Inserting this expression into Eq. \eqref{eq:SlkDef} gives the spherical harmonic power spectrum of flat-spectrum foregrounds
\begin{equation}
\label{eq:fgSlk}
S_\ell^\textrm{fg} (k) = 4 \pi C_\ell^\textrm{fg} \frac{\left[\int_0^\infty \!dr\, r^2 j_\ell (kr) \phi(r) \right]^2}{\int_0^\infty \!dr\, r^2 j_\ell^2 (kr) \phi^2(r)},
\end{equation}
where $C_\ell^\textrm{fg}$ is the angular power spectrum of the foregrounds. For a given survey geometry and foreground model, one can evaluate this expression numerically to derive the signature of foregrounds as manifested in the spherical harmonic power spectrum. Before doing so, however, it is helpful to evaluate $S_\ell^\textrm{fg} (k)$ analytically in various limiting regimes on the $\ell$-$k$ plane to gain intuition for how the spherical harmonic power spectrum behaves. To identify these regimes (which demonstrate qualitatively different behavior), consider Fig. \ref{fig:bessels}, which shows $j_\ell (kr)$ for various choices of $\ell$ and $k$. Not all parts of these curves are relevant to the integrals in Eq. \eqref{eq:fgSlk}, since the radial extent of the survey $\phi(r)$ (indicated by the grey band) picks out only regions where $r \approx r_0$ to integrate over. Roughly speaking, there are two limiting regimes of interest. The first is where $\ell \sim k r_0$. In this regime, the Bessel functions behave like power laws that rise to a peak. The other regime is where $\ell \lesssim k r_0$. There, the Bessel functions are highly oscillatory, and the radial transform of Eq. \eqref{eq:fgTlm} is closely related to a Fourier transform along the line of sight. In principle, there exist modes with $\ell > kr_0$ exist, but as we argued in Section \ref{sec:FiniteVolume}, these modes have very low signal-to-noise, and we will not consider this regime further.

\subsection{Mostly angular modes: $\ell \sim k r_0$}
\label{sec:MostlyAngular}
As discussed previously, the condition that $\ell \sim k r_0$ is synonymous with the statement that fluctuations are almost entirely in the angular direction. In this regime, the spherical Fourier-Bessel functions are not highly oscillatory, and are instead reasonably smooth. They are thus relatively broad compared to $\phi(r)$. To a good approximation, then, $r^2 j_\ell (kr)$ and $r^2 j_\ell^2 (kr)$ may be factored out of the integrals in Eq. \eqref{eq:fgSlk}, evaluating them at $r = r_0$. What remains is
\begin{eqnarray}
\label{eq:FinalSlfgkHighEll}
S_\ell^\textrm{fg} (k) \Bigg{|}_{\ell \gtrsim k r_0} &\approx&  4 \pi C_\ell^\textrm{fg} r_0^2 \frac{\left[\int_0^\infty \!dr\, \phi(r) \right]^2}{\int_0^\infty \!dr\, \phi^2(r)} \nonumber \\
&\sim& 4 \pi C_\ell^\textrm{fg} r_0^2 \Delta r_\textrm{survey},
\end{eqnarray}
where the final approximation is exact only for a survey that has a tophat profile in the radial direction, but still likely to be correct up to a factor of order unity otherwise. One sees that the $k$ dependence of $S_\ell^\textrm{fg} (k)$ drops out, and the measurement is essentially of the angular power spectrum of foregrounds because the radial Bessel transform effectively just averages all the radial fluctuations of the survey together.

\subsection{Mostly radial modes: $\ell \ll k r_0$}
\label{sec:MostlyRadialNoInterferometry}

At low $\ell$ values, most of the spatial variations in one's basis functions are along the line-of-sight. We enter this low $\ell$ regime when $\ell \ll kr_0$, in which case the Bessel functions may be approximated as
\begin{equation}
j_\ell (kr) \approx \frac{1}{kr} \sin \left(kr-\frac{\pi \ell}{2} \right).
\end{equation}
In this limit, the integral in the numerator of Eq. \eqref{eq:fgSlk} becomes
\begin{eqnarray}
&&  \int_0^\infty \!dr\,  r^2 j_\ell (kr) \phi(r) =\frac{1}{k}  \int_0^\infty \! dr\, r \sin \left(kr- \frac{\pi \ell}{2} \right)  \phi(r)  \nonumber \\
 && =- \frac{1}{k^2} \frac{\partial}{\partial \alpha} \left\{ \textrm{Re} \left[ \int_0^\infty \! dr \,e^{-i\alpha kr +i \pi \ell / 2} \phi(r) \right] \right\}_{\alpha = 1},
\end{eqnarray}
where the ``$\alpha = 1$" label signifies that $\alpha$ is to be set to unity after the partial derivative is taken. To proceed, we expand the definition of $\phi(r)$ to include the (unphysical) region of $r < 0$, declaring $\phi(r)$ to be zero when $r < 0$. This allows us to extend the integral to $-\infty$, which enables us to interpret it as a Fourier transform. Further defining $\Phi (r - r_0) \equiv \phi (r)$ to be a re-centered version of the radial profile of the survey for our convenience, we have
\begin{eqnarray}
&&\int_0^\infty \!dr\,  r^2 j_\ell (kr) \phi(r) \nonumber \\
&=& \frac{1}{k} \left[ r_0 \sin \left( k r_0- \frac{\pi \ell}{2} \right) \widetilde{\Phi} (k) - \cos\left( k r_0- \frac{\pi \ell}{2} \right) \widetilde{\Phi}^\prime (k) \right], \qquad
\end{eqnarray}
where the $\widetilde{\Phi}^\prime \equiv \partial \widetilde{\Phi} / \partial k$. Using similar manipulations, the denominator of Eq. \eqref{eq:fgSlk} can be shown to be
\begin{eqnarray}
\label{eq:DenomMostlyRadialFG}
&&\int_0^\infty \!dr\,  r^2 j_\ell^2 (kr) \phi^2(r) \nonumber \\
&=& \frac{1}{2 k^2} \left[ \int_{-\infty}^\infty \!dr\, \Phi^2 (r) - \cos\left(2 k r_0- \pi \ell \right) \widetilde{\Phi} \star\widetilde{\Phi} (2 k)  \right], \qquad
\end{eqnarray}
where $\star$ denotes a convolution. To simplify matters, we may ignore the second term in this expression because it is small compared to the first. To see this, note that the first term can be written as $\widetilde{\Phi^2} (0)$. The relative size of the two terms is therefore determined by the relative magnitudes of $\widetilde{\Phi^2} (0)$ and $ \widetilde{\Phi} \star\widetilde{\Phi} (2 k) $. Now, $\widetilde{\Phi} (k)$ is a function that is reasonably sharply peaked about $k=0$, with a characteristic width given by $\sim 1/ \Delta r_\textrm{survey}$. We expect $\widetilde{\Phi^2}$ to be slightly broader; a back-of-the-envelope estimate would suggest that $\widetilde{\Phi^2}$ is roughly a factor of $\sqrt{2}$ broader than $\widetilde{\Phi}$. Continuing with our approximate line of reasoning, one would then expect $ \widetilde{\Phi} \star\widetilde{\Phi} (2 k)$ to be approximately the same size as $ \widetilde{\Phi} (\sqrt{2} k)$, which is likely to be small because typical $k$ values are of order $\sim 1/ \Delta r_\textrm{survey}$ or larger, placing one beyond the characteristic width of $\widetilde{\Phi}$, where the amplitude is much suppressed compared to the $k=0$ point. We thus conclude that the second term of Eq. \eqref{eq:DenomMostlyRadialFG} may be neglected.

Putting everything together, we obtain
\begin{eqnarray}
\label{eq:OscOsc}
 S_\ell^\textrm{fg} (k) \approx  \frac{8 \pi C_\ell^\textrm{fg}}{\int_{-\infty}^\infty \!dr\, \Phi^2 (r)}  \bigg{[}&& r_0^2 \sin^2 \left( k r_0- \frac{\pi \ell}{2} \right) \widetilde{\Phi}^2 (k) \nonumber \\
&& -r_0 \sin \left( 2 k r_0- \pi \ell \right) \widetilde{\Phi} (k) \widetilde{\Phi}^\prime (k) \nonumber \\
&&+  \cos^2\left( k r_0- \frac{\pi \ell}{2} \right) \widetilde{\Phi}^{\prime 2} (k) \bigg{]}. \qquad
\end{eqnarray}
This result can be further simplified by considering the length scales involved. Recall that that the key approximation of this subsection is that the spatial fluctuations are mostly along the radial direction. For a survey with radial resolution $\Delta r_\textrm{res}$ (determined by an instrument's spectral resolution), a natural choice for a bin size in $k$ would be $\sim\!2 \pi / \Delta r_\textrm{res}$. Since the value of $k$ is multiplied by $r_0$ inside the oscillatory terms of Eq. \eqref{eq:OscOsc}, and $r_0 \gg \Delta r_\textrm{res}$, it follows that one goes through many cycles of the sinusoids within each bin in $k$ in any practical measurement. The middle term of Eq. \eqref{eq:OscOsc} thus averages to zero, while the squared sinusoids average to $1/2$. We thus have
\begin{equation}
S_\ell^\textrm{fg} (k) \approx 4 \pi \frac{ C_\ell^\textrm{fg}}{\int_{-\infty}^\infty \!dr\, \Phi^2 (r)}  \left[ r_0^2  \widetilde{\Phi}^2 (k) +\widetilde{\Phi}^{\prime 2} (k)  \right]
\end{equation}
Now, the two terms seen here that comprise $S_\ell^\textrm{fg} (k)  $ are not of equal importance. Dimensional analysis suggests that the derivative of $\Phi$ is of order $\Phi^\prime \sim \Phi / \Delta r_\textrm{survey}$, while the derivative of its Fourier transform $\widetilde{\Phi}$ is of order $\widetilde{\Phi}^\prime \sim \widetilde{\Phi} \Delta r_\textrm{survey}$, a fact that can be verified by testing various functional forms for $\Phi$. The first term in our expression for $S_\ell^\textrm{fg} (k)$ is thus larger than the second term by a factor of $(r_0 / \Delta r_\textrm{survey})^2$, which greatly exceeds unity for high-redshift measurements. These simplifications yield the final expression
\begin{equation}
\label{eq:FinalSlfgkLowEll}
S_\ell^\textrm{fg} (k) \Bigg{|}_{\ell \lesssim k r_0} \approx 4 \pi  C_\ell^\textrm{fg}   \frac{r_0^2  \widetilde{\Phi}^2 (k)}{\int_{-\infty}^\infty \!dr\, \Phi^2 (r)}. 
\end{equation}
This result is essentially identical to its flat-sky counterpart on the $k_\perp$-$k_\parallel$ plane. There, the foregrounds were seen to be confined mostly to low $k_\parallel$ values, with the characteristic width of the fall-off towards higher $k_\parallel$ of $\sim 1 / \Delta r_\textrm{survey}$, as expected from the Fourier transform of data that spans a length of $\Delta r_\textrm{survey}$. Here, in the regime where our modes are dominated by radial fluctuations, we have $k$ taking the place of $k_\parallel$. But the behavior is the same, since $\widetilde{\Phi} (k)$ falls off as $\sim 1 / \Delta r_\textrm{survey}$.

\subsection{Numerical Results}
\label{sec:Numerics}
Summarizing the last two results, it is pleasing to note that the even though Eqs. \eqref{eq:FinalSlfgkHighEll} and \eqref{eq:FinalSlfgkLowEll} were derived as different limiting cases, the latter converges to the former when $k\rightarrow 0$. This suggests a rather smooth transition between the two regimes and a simple signature of foregrounds as a function of $\ell$ and $k$: at low $k$, the foregrounds are a strong contaminant, but their influence quickly falls off towards higher $k$.

We confirm this behavior in the top right panel of Figure \ref{fig:fgSigs} by plotting a numerically computed $S_\ell^\textrm{fg} (k)$. The survey parameters are assumed to be the same as in Section \ref{sec:Foregrounds}. There is a qualitative similarity between the flat-sky plot of $P(k_\perp, k_\parallel)$ in the top left panel, and the curved-sky plot of $S_\ell (k)$ in the top right. This suggests that the latter will be just as successful as the former in localizing foregrounds in their respective planes. Quantitatively, one sees a sharp drop-off towards higher $k$ (or $k_\parallel$), with some ringing due to our cosine radial profile. Admittedly, the drop-off is not quite as steep as one might hope, given that the foregrounds can easily be six to nine orders of magnitude brighter than the cosmological in power spectrum units \citep{santos_et_al2005,jelic_et_al2008,bernardi_et_al2009,bernardi_et_al2010}. However, a large number of tools can be employed to further suppress foregrounds at high $k$ (or $k_\parallel$). For example, foregrounds can be filtered or directly subtracted, whether via the construction of foreground models or through blind methods \citep{wang_et_al2006,gleser_et_al2008,liu_et_al2009a,bowman_et_al2009,liu_et_al2009b,harker_et_al2009,petrovic_and_oh2011,paciga_et_al2011,Parsons_et_al2012b,liu_and_tegmark2012,chapman_et_al2012,chapman_et_al2013,wolz_et_al2014,shaw_et_al2014a,shaw_et_al2014b,wolz_et_al2015}. Leakage of foregrounds from low $k$ to high $k$ can be mitigated by imposing tapering functions to apodize the radial profile $\phi(r)$ \citep{Thyagarajan2013}. This would, for instance, reduce the Fourier space ringing from the cosine form of Eq. \eqref{eq:CosineRadial}, which causes the horizontal stripes that are visually obvious in the top row of Figure \ref{fig:fgSigs}. Finally, statistical methods can be employed to selectively downweight foreground contaminated modes, whether prior to the squaring of temperature data in power spectrum estimation \citep{liu_and_tegmark2011,liu_et_al2014a,trott_et_al2016} or after \citep{dillon_et_al2014,liu_et_al2014b}. Our goal here was only to show that $S_\ell (k)$ is just as viable a foreground diagnostic for the curved sky as $P(k_\perp, k_\parallel)$ is for the flat sky, and Figure \ref{fig:fgSigs} shows that this is indeed the case.

\section{Interloper lines in the spherical harmonic power spectrum}
\label{sec:Interlopers}
Aside from broadband foregrounds that are spectrally smooth, some intensity mapping surveys must also deal with the problem of interloper lines, where emission from two different spectral lines that are sourced at different radial distances may nonetheless redshift into the same observing band. More concretely, an interloper line with a rest frequency of $\nu_\textrm{rest}^\prime$ emitted at redshift $z^\prime$ will appear at the same observed frequency as another line (say, the one targeted by an intensity mapping survey) with rest frequency $\nu_\textrm{rest}$ at redshift $z$ if $(1+z^\prime)  / \nu_\textrm{rest}^\prime = (1+z)  / \nu_\textrm{rest}$. The interloper line thus acts as an additional foreground contaminant. For $21\,\textrm{cm}$ intensity mapping this is typically not a problem, simply because there lack plausible spectral line candidates with appropriate rest frequencies. In contrast, [CII] and CO lines are both candidates for intensity mapping surveys, and can easily be confused with one another.

Since interloper lines may themselves trace cosmic structure (albeit at different redshifts), they are not spectrally smooth foreground contaminants, and thus cannot be mitigated by the methods described in the rest of this paper. To deal with this, a variety of techniques have been proposed in the literature, including source masking \citep{silva_et_al2015,yue_et_al2015,breysse_et_al2015}, cross-correlation with external datasets \citep{visbal_and_loeb2010,gong_et_al2012,gong_et_al2014}, comparison to companion lines \citep{kogut_et_al2015}, and the exploitation of angular fluctuations to reconstruct three-dimensional source distributions \citep{dePutter_et_al2014}. Recently, \citet{cheng_et_al2016} and \citet{lidz_and_taylor2016} proposed a method for separating interloper lines by invoking the statistical isotropy of the cosmological signal. The key observation is that the rest frequency of a line enters the frequency-radial distance mapping of Eq. \eqref{eq:ComovingDistDef} in a different way than it does in the angle-transverse distance conversion of Eq. \eqref{eq:AngularConversion}. If emission from an interloper line is mistaken as the targeted line in a survey, it will be mapped to incorrect cosmological coordinates. As a result, the emission will no longer be statistically isotropic, in contrast to emission from the targeted line, which will have been mapped correctly and thus will be statistically isotropic. In terms of the power spectrum, emission from the targeted line will appear in the cylindrical power spectrum $P(k_\perp, k_\parallel)$ as a function of $k \equiv (k_\perp^2 + k_\parallel^2)^{1/2}$ only, while interloper emission will have a non-trivial dependence on $k_\perp$ and $k_\parallel$. This difference in $k_\perp$-$k_\parallel$ signature provides a way to identify interloper emission.

In this section, we build on the work of \citet{cheng_et_al2016} and \citet{lidz_and_taylor2016}, generalizing their flat-sky treatment to the curved sky using the spherical harmonic power spectrum. Our goal will be to show that just as $P(k_\perp, k_\parallel)$ is no longer just a function of $k$ if the incorrect rest frequency $\nu_\textrm{inc}$ is assumed, $S_\ell (k)$ will similarly develop a dependence on $\ell$ under those circumstances. To begin, we note that Eq. \eqref{eq:SHTdef} is always exact, since it only relies on angular information, which does not require knowledge of the rest frequency of the spectral line. The assumption of an incorrect rest frequency enters only in Eq. \eqref{eq:FBdef}, when one must map frequencies to radial distances. Suppose some emission originates from a comoving location $\mathbf{r} = r \rhat$. If the incorrect frequency-radial distance relation is used due to a mistaken assumption about the rest frequency of the emission, this emission will be mapped to a location $\mathbf{r} \equiv s(r) \rhat$ instead, where $s$ is the incorrect radial distance, which is a function of the correct distance $r$. As a result, Eq. \eqref{eq:FBdef} becomes $\overline{T}_{\ell m}^\textrm{inc} (k)$, the incorrectly mapped version of $\overline{T}_{\ell m} (k)$, and take the form
\begin{equation}
\overline{T}_{\ell m}^\textrm{inc} (k) \equiv \sqrt{\frac{2}{\pi}} \int d^3 r j_\ell (kr) Y_{\ell m}^* (\rhat) T[s(r) \rhat] \phi[s(r) \rhat],
\end{equation}
where we have included the finite volume of our survey via the function $\phi$, just as we did in the previous section. Writing the $T\phi$ term in terms of their Fourier transforms and repeating steps analogous to the ones used between Eqs. \eqref{eq:YetAnotherTellm} and \eqref{eq:TlmTkConversion}, we obtain
\begin{eqnarray}
\overline{T}_{\ell m}^\textrm{inc} (k) = i^\ell 4 \sqrt{2 \pi}  \int \frac{d^3 k^\prime}{(2\pi)^3} \frac{d^3 k^{\prime \prime}}{(2\pi)^3} Y_{\ell m} (\khat^\prime ) \widetilde{\phi} (\mathbf{k}^\prime - \mathbf{k}^{\prime \prime}) \nonumber \\
\times \widetilde{T} (\mathbf{k}^{\prime \prime}) \int dr r^2 j_\ell (kr) j_\ell [ k^\prime s(r)]. \quad 
\end{eqnarray}
To relate this to the power spectrum, we square this expression, take the ensemble average, and average over $m$ values. Performing manipulations similar to those that led to Eq. \eqref{eq:TotallyUnsold} results in
\begin{eqnarray}
&&\frac{\sum_{m = -\ell}^\ell\langle | \overline{T}_{\ell m}^\textrm{meas} (k) |^2 \rangle}{2\ell + 1} \approx P(\overline{k}) \frac{2}{\pi^4} \int d^3 r \phi^2 (\mathbf{r}) \nonumber \\
&&\times \left( \int dr^\prime r^{\prime 2} d k^\prime k^{\prime 2} j_\ell (k^\prime r)   j_\ell(k r^\prime) j_\ell [ k^\prime s (r^\prime)] \right)^2, \qquad
\end{eqnarray}
where $\overline{k}$ is some wavenumber that is not necessarily equal to $k$. In other words, with an incorrect mapping of radial distances, we should not necessarily expect $\langle | \overline{T}_{\ell m}^\textrm{meas} (k) |^2 \rangle$ to probe a distribution of power that is sharply peaked around $k$. Any bias in the probed wavenumber, however, is irrelevant for our present purposes, which is simply to show that an $\ell$ dependence is acquired in our (no longer isotropic) estimate of the power spectrum. Performing the $k^\prime$ integral using Eq. \eqref{eq:BesselOrthog} (but with $r$ and $k$ swapping roles) and inserting the result into Eq. \eqref{eq:SlkDef}, one obtains
\begin{eqnarray}
S_\ell^\textrm{inc} (k)=&&P(\overline{k}) \left[ \int d^3 r \phi^2 (\mathbf{r}) j_\ell^2 (kr) \right]^{-1} \nonumber \\
&&\times \int d^3 r \frac{\phi^2 (\mathbf{r})}{s^\prime [s^{-1} (r)]} \left(\frac{s^{-1} (r)}{r}\right)^2 j_\ell^2 [ks^{-1} (r)] \qquad
\end{eqnarray}
for the estimated spherical harmonic power spectrum under the assumption of a mistaken rest frequency. Here, $s^\prime \equiv \partial s / \partial r$ (i.e., the derivative of the incorrectly mapped radial distance with respect to the true radial distance) and $s^{-1}$ denotes an inverse mapping, not a reciprocal. Notice that if the rest frequency is correct (i.e., one is dealing with emission from the targeted line rather than the interloper line), then $s$ is the identity function, $s^\prime$ is unity, and the two integrals cancel to leave a result that is $\ell$-independent. In general, however, the result will be $\ell$-dependent. We thus conclude that just as anisotropies in $P(k_\perp, k_\parallel)$ can be used to detect interloper lines within the flat-sky approximation, $S_\ell (k)$ can be used in the same way for a full curved-sky treatment.

\section{Spherical Harmonic Power Spectrum Measurements with Interferometers}
\label{sec:Interferometry}

In previous sections, we have focused on understanding the \emph{intrinsic} spherical harmonic power spectrum $S_\ell (k)$ without the inclusion of any instrumental effects other than a selection function to account for survey geometry. For some intensity mapping efforts, the exclusion of these effects will not result in major differences in $S_\ell (k)$. For instance, at higher frequencies (say, those relevant to [CII] intensity mapping) it is common to perform intensity mapping with traditional single dish telescopes and spectrometers. With such systems, the equations derived so far in this paper are a reasonable approximation for what one might see in real data, perhaps with the addition of a high noise component at high $\ell$ and $k$ to reflect finite angular and spectral resolution. In contrast, at low frequencies it is common to perform intensity mapping using radio interferometers. In this section, we will show that with data from interferometers, $S_\ell (k)$ behaves qualitatively differently from what we have considered so far. Despite these differences, once the data (and any accompanying metrics for describing their statistical properties) are reduced to modes in the spherical Fourier-Bessel basis, it is irrelevant whether they were collecting using single dish telescopes or interferometers. The spherical Fourier-Bessel basis and the spherical harmonic power spectrum $S_\ell (k)$ may thus be a useful meeting point for cross-correlations between the $21\,\textrm{cm}$ and CO/[CII] lines (e.g., as proposed in \citealt{lidz_et_al2011}).

Interferometers are frequently used for intensity mapping measurements because they are essentially Fourier-space instruments, with each baseline of an interferometer directly sampling a fringe pattern that approximates one of the spatial Fourier modes of interest. They are therefore a relatively inexpensive way to perform high-sensitivity measurements of the power spectrum. However, the picture of an interferometer as a Fourier-space instrument is precisely correct only in the limit that the sky is flat. This assumption is typically invoked in derivations of estimators for connecting interferometric measurements to power spectra \citep{hobson_et_al1995,white_et_al1999,padin_et_al2001,halverson_et_al2002,hobson_et_al2002,myers_et_al2003,parsons_et_al2012a,parsons_et_al2014}. It is, however, explicitly violated by the wide-field nature of many instruments built for intensity mapping. In this section, we will address this shortcoming, using the spherical Fourier-Bessel formalism to relate interferometric data to the cosmological power spectrum in a way that fully respects curved sky effects.

For the purposes of three-dimensional intensity mapping experiments, interferometers come with the added complication of being inherently chromatic instruments. Consider, for example, the visibility measured by a single baseline of an interferometric array:
\begin{subequations}
\begin{eqnarray}
V(\mathbf{b}, \nu) &=& \int d\Omega \phi(r_\nu) A(\rhat, \nu) I(\rhat, \nu)e^{ - i 2\pi \nu  \mathbf{b} \cdot\rhat / c} \label{eq:VisDef}\\
&\equiv&\int d \Omega  \phi(r_\nu)  B(\rhat, \nu) T(\rhat, \nu) e^{ - i 2\pi \nu  \mathbf{b} \cdot\rhat / c} \label{eq:CurvedVisibility}\\
&\approx& \int \frac{d^2 r_\perp}{r_\nu^2} \phi(r_\nu) B(\rhat, \nu) T(\rhat, \nu)e^{ - i 2\pi \nu  \mathbf{b} \cdot \mathbf{r}_\perp / c r_\nu}, \label{eq:FlatVisibility}\qquad\quad 
\end{eqnarray}
\end{subequations}
where in the last line we invoked the narrow-field, flat-sky approximation, allowing a ``line-of-sight" direction to be unambiguously identified and a position vector $\mathbf{r}_\perp$ transverse to this direction to be defined. In the penultimate line we used the Rayleigh-Jeans Law to convert from intensity to brightness temperature, defining a modified primary beam
\begin{equation}
\label{eq:Bdef}
B(\rhat, \nu) \equiv \frac{2 k_B}{c^2} \nu^2 A(\rhat, \nu).
\end{equation}
One sees that in the flat-sky limit, the complex exponential takes the form of $\exp \left( - i \mathbf{k}_\perp \cdot \mathbf{r}_\perp \right)$, and thus the baseline probes a spatial mode perpendicular to the line of sight with wavevector $\mathbf{k}_\perp = 2 \pi \nu \mathbf{b} / c r_\nu$. The key feature to note here is that this spatial scale is dependent on $\nu$. Interferometers are therefore inherently chromatic in the sense that the Fourier mode probed by a particular baseline depends on frequency, particularly if the baseline is long. This complicates the power spectrum measurement, for in order to access Fourier modes along the line of sight (characterized by wavenumber $k_\parallel$), it is necessary to perform a Fourier transform along the frequency axis. At least for data from a single baseline, the chromaticity means that $\mathbf{k}_\perp$ is not held constant during the line of sight Fourier transform. This causes couplings between $k_\parallel$ and $\mathbf{k}_\perp$ modes, and is responsible for the wedge feature that has been discussed extensively in the previous literature. The wedge arises when the chromaticity of an interferometer imprints this chromaticity on observed foregrounds. Being spectrally smooth, the foregrounds should in principle be localized to low $k_\parallel$ modes (as we saw in the top panels of Figure \ref{fig:fgSigs}), but in practice the imprinted chromaticity causes them to appear at higher $k_\parallel$ modes in a wedge-like signature.

The wedge is both a problem and an opportunity. The wedge is a problem because it increases (compared to a theoretically ideal situation with no instrument chromaticity) the number of Fourier modes that are foreground-dominated and thus unavailable for a measurement of the cosmological signal. These unavailable modes are often the ones that are highest in signal-to-noise, resulting in a significant reduction in sensitivity \citep{pober_et_al2014,chapman_et_al2016}. However, the wedge is also an opportunity because it can be shown (in a reasonably general manner) that it is limited in extent, i.e., the foreground contamination does not extend beyond the confines of the wedge shape. Observations can therefore be targeted at modes that are outside the wedge, and instruments may be designed conservatively to optimize such observations \citep{parsons_et_al2012a}. Indeed, this is the general principle behind the design of HERA \citep{deboer_et_al2016}.


That smooth spectrum foregrounds have a well-defined signature in the form of the wedge is one of the reasons that recent works have espoused the $P(k_\perp,k_\parallel)$ power spectrum as a useful diagnostic for data analysis. In order for our proposed statistic $S_\ell (k)$ to be useful in the same way, it is necessary to show that the chromatic influence of an interferometer also gives a well-defined and well-localized signature $\ell$-$k$ space. We will do so in the following subsections once we have established the connection between curved sky power spectra and interferometeric data, finding that foregrounds are again localized to a wedge. We will focus on single-baselines analyses of the data, as this provides a conservative estimate for the extent of the foreground wedge in $S_\ell (k)$. Multi-baseline information can be used to alleviate wedge effects, because one can essentially combine data from different frequencies and different baselines that have the same ratio $\nu \mathbf{b} /  r_\nu$, alleviating the chromatic effects that caused the wedge in the first place. There thus exist methods for reducing the extent of the wedge, and our single-baseline treatment should be considered a worst-case scenario.

\subsection{Delay spectrum power spectrum estimation}
\label{sec:DelayIntro}
To estimate the power spectrum from a single baseline, one begins by forming the \emph{delay spectrum} of the baseline's visibility. This is accomplished by Fourier transforming the visibility along the frequency axis to obtain
\begin{equation}
\label{eq:DelayDef}
\widetilde{V} (\mathbf{b}, \tau) \equiv \int \!d\nu\, \gamma(\nu) e^{i 2\pi \nu \tau} V(\mathbf{b}, \nu),
\end{equation}
where $\gamma (\nu)$ is an optional tapering function chosen by the data analyst. Given that $V$ approximates the sky brightness Fourier transformed in the axes perpendicular to the line of sight, $\widetilde{V}$ serves as an approximation for the $\widetilde{T}(\mathbf{k})$. The delay spectrum can then be squared and normalized to yield an estimator for the power spectrum $P(k)$.

As we discussed above, a single baseline probes different $\mathbf{k}_\perp$ scales at different frequencies. Power spectra estimated using delay spectra are therefore often considered mere approximations to ``true" power spectra. However, an estimator formed from the delay spectrum represents a perfectly valid estimator, so long as error statistics are included in the final results. The quoted error statistics on a power spectrum estimate $\widehat{P}(k_*)$ at some spatial scale $k_*$ should include not only the error bars on the value of $\widehat{P}$ itself, but also window functions for describing the (sometimes broad) distribution of $k$ values that contribute to a power estimate that is centered on $k = k_*$. Because single-baseline estimators have a chromatic scale-dependence, their resulting window functions will be wider than what might be in principle achievable using a well-controlled multi-baseline approach. In general, however, the latter will still give windows of non-zero width (due to a combination of finite-volume and analysis pipeline effects), and in that sense a delay spectrum power spectrum with well-documented error statistics is not any more of an approximation than any other method.\footnote{The term ``window function" is unfortunately rather overused. In various parts of the literature, it has been used to refer to what we have called the tapering function $\gamma$ in this paper, and in other parts of the literature it has been used to describe what we have called the survey profile $\phi$. In this paper, a window function will \emph{always} refer to the function that describes the linear combination of true power spectrum probed by one's statistical estimator of the power spectrum. A mathematically precise definition for the window functions of our particular estimator will be provided in Eqs. \eqref{eq:delayWindow} and \eqref{eq:DelayWindowFcts}.}

In the following subsections we establish the framework for single-baseline analyses of the power spectrum in the curved sky. Section \ref{sec:WindowFcts} computes the window functions associated with delay spectrum power spectrum estimation. Section \ref{sec:SingleBlNorm} provides a rigorous derivation of power spectrum normalization, using our spherical harmonic formalism to incorporate curved sky treatments that have so far been neglected in the literature. Section \ref{sec:CurvedSkyWedge} then demonstrates how the foreground wedge signature seen in $P(k_\perp, k_\parallel)$ spectra is preserved in $S_\ell (k)$.

\subsection{Window functions of a delay-based power spectrum estimate}
\label{sec:WindowFcts}

As mentioned above, one estimates the power spectrum from a single baseline by first forming the delay spectrum $\widetilde{V}$, followed by a subsequent squaring of the result. Computing the window functions of such an estimate requires relating our measurements to a theoretical power spectrum. To do so, we take the definition of a single baseline's visibility from Eq. \eqref{eq:CurvedVisibility} and expand the temperature field in spherical harmonics, giving
\begin{equation}
V(\mathbf{b}, \nu) = \sum_{\ell m} \phi (r_\nu) a_{\ell m} (\nu) f_{\ell m} (\mathbf{b}, \nu),
\end{equation}
where we have defined
\begin{equation}
\label{eq:flm}
f_{\ell m}(\mathbf{b}, \nu) \equiv \int d \Omega  B(\rhat, \nu)  Y_{\ell m} (\rhat) e^{ - i 2\pi \nu  \mathbf{b} \cdot\rhat / c}.
\end{equation}
as the response of a baseline $\mathbf{b}$ to an excitation of the spherical harmonic with indices $\ell$ and $m$. The detailed properties of this response function have previously been explored in the literature \citep{shaw_et_al2014a,zheng_et_al2014,shaw_et_al2014b,zhang_et_al2016a,zhang_et_al2016b}. 
Here, we relate this response function to a delay spectrum approach. To proceed, we use Eq. \eqref{eq:FBdef} (or rather, the inverse of the transformation it describes) to express $a_{\ell m}$ in terms of its spherical Fourier-Bessel expansion, giving
\begin{equation}
\label{eq:VisTlmConnection}
V(\mathbf{b}, \nu) = \sqrt{\frac{2}{\pi}} \sum_{\ell m} \int \!dk\, k^2 j_\ell (k r_\nu)  \overline{T}_{\ell m} (k) f_{\ell m} (\mathbf{b}, \nu) \phi(r_\nu).
\end{equation}
Forming the delay spectrum $\widetilde{V}$ from this then yields
\begin{equation}
\label{eq:VtildeInToverline}
\widetilde{V} (\mathbf{b}, \tau) =  \sqrt{\frac{2}{\pi}} \sum_{\ell m} \int \!dk\, k^2 g_{\ell m} (k; \mathbf{b}, \tau) \overline{T}_{\ell m} (k),
\end{equation}
where
\begin{equation}
\label{eq:glm}
g_{\ell m}(k; \mathbf{b}, \tau) \equiv \int d\nu e^{i 2\pi \nu \tau} j_\ell (k r_\nu) f_{\ell m} (\mathbf{b}, \nu) \phi(r_\nu) \gamma (\nu).
\end{equation}

Now suppose the measured sky consists only of the cosmological signal. The $\overline{T}_{\ell m} (k)$ modes are then directly related to the power spectrum via Eq. \eqref{eq:CurvedPspecDef}, and the ensemble average of the square of the delay spectrum reduces to
\begin{eqnarray}
\label{eq:delayWindow}
\langle |\widetilde{V} (\mathbf{b}, \tau) |^2 \rangle && = \frac{2}{\pi} \sum_{\ell m \ell^\prime m^\prime} \int dk dk^\prime k^2 k^{\prime 2} \langle \overline{T}_{\ell m} (k) \overline{T}_{\ell^\prime m^\prime}^* (k^\prime)\rangle\nonumber \\
&& \qquad \times \, g_{\ell m} (k; \mathbf{b}, \tau) g_{\ell^\prime m^\prime}^* (k^\prime; \mathbf{b}, \tau)  \nonumber \\
&& = \sum_{\ell} \int dk W_\ell^\textrm{unnorm} (k; \mathbf{b}, \tau) P(k), \qquad 
\end{eqnarray}
where
\begin{equation}
\label{eq:DelayWindowFcts}
W_\ell^\textrm{unnorm} (k; \mathbf{b}, \tau) \equiv \frac{2k^2}{\pi} \sum_m | g_{\ell m} (k ; \mathbf{b}, \tau)|^2
\end{equation}
are the (unnormalized) window functions. For given values of $\mathbf{b}$ and $\tau$, Eq. \eqref{eq:delayWindow} shows that the window function describes the linear combination of modes on the $\ell$-$k$ plane that are probed by the quantity $\langle |\widetilde{V} (\mathbf{b}, \tau) |^2 \rangle$. If $ |\widetilde{V} (\mathbf{b}, \tau)|^2$ is to be a good estimator of the power spectrum, the window function for each $(\mathbf{b},\tau)$ pair should satisfy two conditions. First, each window function should be reasonably sharply peaked around some location on the $\ell$-$k$, giving a precise measurement of the power spectrum on some scale rather than a broad combination of scales. Second, the window functions for different values of $(\mathbf{b},\tau)$ should be centered on different locations on the $\ell$-$k$ plane. In other words, the ideal collection of window functions should divide the $\ell$-$k$ plane into a set of mutually exclusive and collectively exhaustive cells \citep{tegmark_et_al1998}.

\begin{figure}[!]
	\centering
	\includegraphics[width=0.48\textwidth]{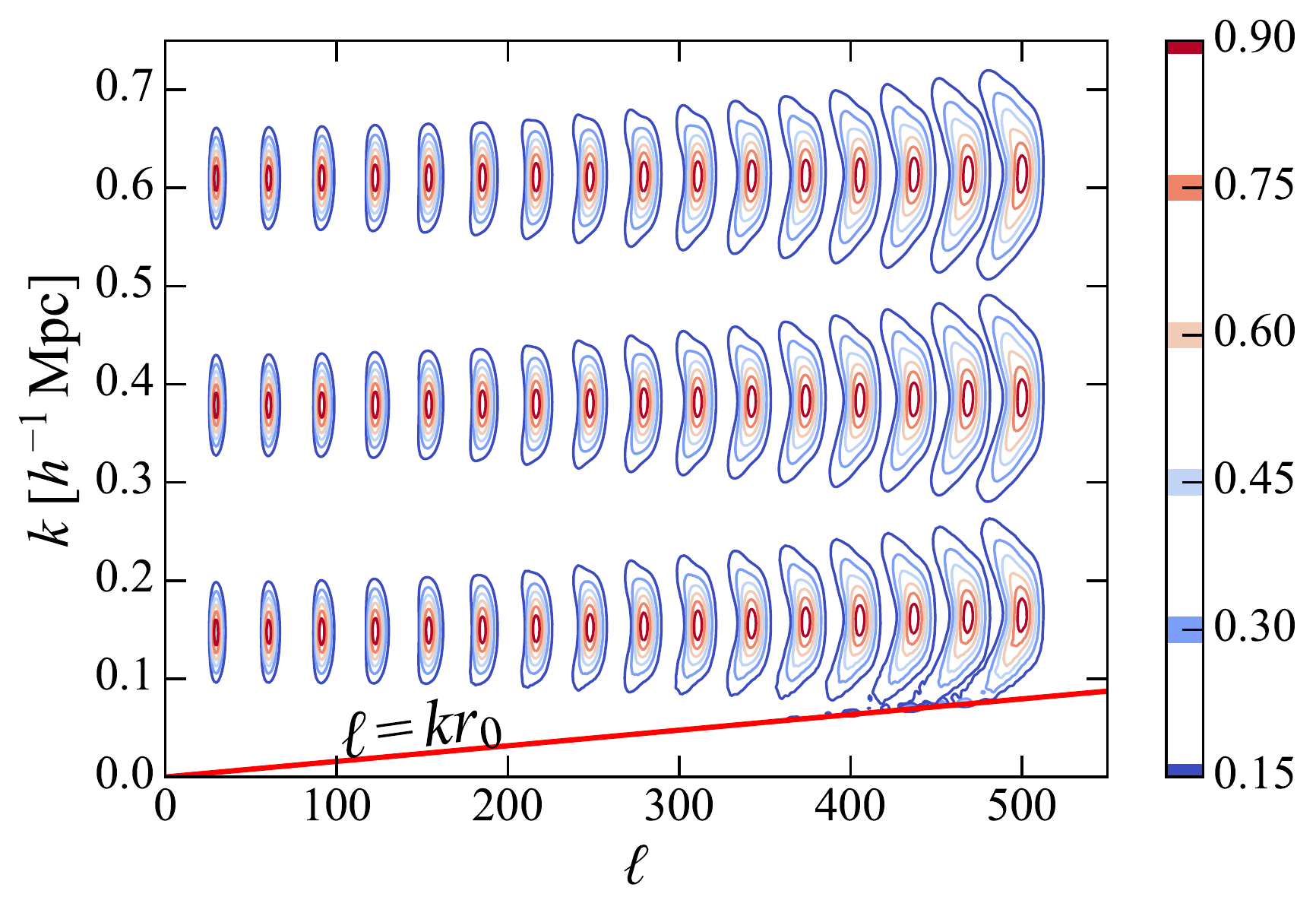}
	\caption{Example window functions on the $\ell$-$k$ plane, given by Eq. \eqref{eq:DelayWindowFcts}. Each set of contours describes the linear combination of modes on the $\ell$-$k$ plane sampled by a power spectrum estimator formed by a particular baseline-delay combination. From bottom to top, the three rows correspond to the windows for estimators with delay $\tau = 273\,\textrm{ns}$, $703\,\textrm{ns}$, and $1133\,\textrm{ns}$. From left to right, each column corresponds to windows for estimators with baseline lengths from $10\,\textrm{m}$ to $190\,\textrm{m}$ in $10\,\textrm{m}$ increments. To allow everything to be easily visualized on a common color scale, each window function is normalized to peak at unity. The boundary $\ell = k r_0$ is demarcated by the bold red line. Parts of the plane below this line are difficult to access, and all window functions are seen to taper off towards the line.}
	\label{fig:windowFcts}
\end{figure}

In Figure \ref{fig:windowFcts} we show example $\ell$-$k$ plane window functions for various choices of $(\mathbf{b},\tau)$, computed using the same survey parameters as in Section \ref{sec:Numerics}. All the window functions tend to taper off towards the line $\ell = kr_0$, consistent with our previous argument that regions below this line are difficult to probe with any substantial signal-to-noise. We find that to a good approximation, the peaks of the window functions are located at
\begin{equation}
\label{eq:PeakLocs}
k \approx 2 \pi \sqrt{\left(\frac{\tau}{\alpha_0}\right)^2 + \left(\frac{b \nu_0}{c r_0}\right)^2}; \quad \ell \approx \frac{2 \pi b \nu_0}{c},
\end{equation}
where $\alpha_0$ is the radial distance-frequency conversion from Eq. \eqref{eq:AlphaConversion} evaluated with the reference frequency set to $\nu_0$, the frequency at the middle of our observational band. These expressions are what one would write down assuming a flat-sky mapping between interferometer parameters $(\mathbf{b}, \tau)$ and spatial fluctuation wavenumbers $\ell$ and $k$. Given this, it is unsurprising that the accuracy of these approximations goes down at low $\ell$, where curved sky effects are expected to be the most important. Nonetheless, the accuracy is reasonable throughout: we find that the $\ell$ location of the peaks predicted by Eq. \eqref{eq:PeakLocs} to be good to $\sim 10\%$ at $\ell \sim 30$, improving to $5\%$ by $\ell \sim 50$ and with further improvements as $\ell$ increases. Nowhere in the $\ell$-$k$ range bracketed by the window functions shown in Figure \ref{fig:windowFcts} do we find the errors to be larger than $10\%$. Our prediction for the $k$ location of the peaks is better yet, with the errors never exceeding $\sim 5\%$, and more typically at the sub-percent level. In any case, our approximations are meant for illustration purposes only. In a practical estimation of power spectra, one should compute the exact window functions (as we have done here by numerical means), and these window functions should accompany any power spectrum results that are presented.

For $|\widetilde{V} (\mathbf{b}, \tau)|^2$ to serve as a useful estimator of the power spectrum, its window functions must not only be centered on different parts of the $\ell$-$k$ plane for different values of $\mathbf{b}$ and $\tau$ (as we have just shown). The windows must also be relatively compact, and we see in Figure \ref{fig:windowFcts} that this is indeed the case. A key feature, however, is that the window functions become elongated in the $k$ direction as one moves to higher $\ell$. This effect is exactly analogous to the $k_\parallel$ elongation of window functions at high $k_\perp$ in the flat-sky case examined in \citet{liu_et_al2014a}, and is due to the fact that the higher $\ell$ (or $k_\perp$) are probed by longer baselines, which (as we discussed in Section \ref{sec:Interferometry}) exhibit a more chromatic response. The elongation seen here is our first hint of the foreground wedge, since an extended window function in $k$ (or $k_\parallel$) will pick up more foreground contamination from the lower portions of the $\ell$-$k$, where foregrounds intrinsically reside. This causes foregrounds to leak ``upwards" on the plane, with the extent of the leakage tracking the increasingly exaggerated elongation towards higher $\ell$ (or $k_\perp$), thus resulting in a wedge-like feature. We will derive the $\ell$-$k$ plane foreground wedge more rigorously in Section \ref{sec:CurvedSkyWedge}. For now, it suffices to say that since the window functions seen in Figure \ref{fig:windowFcts} are reasonably compact, we have successfully demonstrated that $|\widetilde{V} (\mathbf{b}, \tau)|^2$ is just as potent an estimator of the power spectrum in our full curved-sky formalism as it is in the flat-sky.

\subsection{Normalizing a delay-based power spectrum estimate}
\label{sec:SingleBlNorm}

In the previous subsection, we showed that the $|\widetilde{V} (\mathbf{b}, \tau)|^2$ is a suitable estimator for the cosmological power spectrum. However, it is not yet properly normalized. Here, we derive the normalization factor that $|\widetilde{V} (\mathbf{b}, \tau)|^2$ must be divided by to obtain an unbiased estimate of the power spectrum.

From Eq. \eqref{eq:delayWindow}, we see that $\langle |\widetilde{V} (\mathbf{b}, \tau) |^2 \rangle$ measures a weighted sum/integral of the power spectrum. For our estimator to be properly normalized, the weighted sum/integral ought to be a weighted average instead. We can accomplish this by dividing $\langle |\widetilde{V} (\mathbf{b}, \tau) |^2 \rangle$ by the sum/integral of $W_\ell^\textrm{unnorm} (k; \mathbf{b}, \tau)$, which serves as a normalization factor. This normalization can be considerably simplified:
%
\begin{eqnarray}
\label{eq:DelayNormSimplification}
 \sum_{\ell m} \!\!\!&&\int dk W_\ell^\textrm{unnorm} (k; \mathbf{b}, \tau) \nonumber \\
 &=&  \frac{2}{\pi}  \sum_{\ell m} \!\int dk \,k^2 |g_{\ell m} (k; \mathbf{b}, \tau)|^2 \nonumber \\
&=& \frac{2}{\pi} \int d\nu d\nu^\prime e^{i 2\pi (\nu-\nu^\prime) \tau} \int dk k^2 j_\ell (kr_\nu) j_\ell(k r_{\nu^\prime})  \nonumber \\
&& \qquad  \phi(r_\nu) \phi(r_{\nu^\prime}) \gamma (\nu) \gamma (\nu^\prime) \sum_{\ell m}   f_{\ell m} (\mathbf{b}, \nu) f_{\ell m}^* (\mathbf{b}, \nu^\prime) \qquad \nonumber \\
&=& \int \frac{d\nu}{r_\nu} \frac{d\nu^\prime}{r_{\nu^\prime}} e^{i 2\pi (\nu-\nu^\prime) \tau} \delta^D (r_\nu - r_{\nu^\prime}) \nonumber \\
&& \qquad  \phi(r_\nu) \phi(r_{\nu^\prime}) \gamma (\nu) \gamma (\nu^\prime) \sum_{\ell m}   f_{\ell m} (\mathbf{b}, \nu) f_{\ell m}^* (\mathbf{b}, \nu^\prime), \qquad
\end{eqnarray}
where in the last line we invoked the orthnormality of spherical bessel functions with different arguments. Continuing, we have
\begin{eqnarray}
&& \int \frac{d\nu}{r_\nu} \frac{d\nu^\prime}{r_{\nu^\prime}} e^{i 2\pi (\nu-\nu^\prime) \tau} \frac{\delta^D (\nu - \nu^\prime)}{\alpha_\nu} \nonumber \\
&& \qquad  \phi(r_\nu) \phi(r_{\nu^\prime}) \gamma (\nu) \gamma (\nu^\prime) \sum_{\ell m}   f_{\ell m} (\mathbf{b}, \nu) f_{\ell m}^* (\mathbf{b}, \nu^\prime)  \nonumber \\
&=& \int \frac{d\nu}{r_\nu^2 \alpha_\nu} \phi^2(r_\nu) \gamma^2 (\nu) \sum_{\ell m}  | f_{\ell m} (\mathbf{b}, \nu)|^2 \nonumber \\
&=& \int d\Omega d\nu \frac{B^2 (\rhat, \nu) \phi^2(r_\nu) \gamma^2 (\nu)}{r_\nu^2 \alpha_\nu},
 \end{eqnarray}
 where in the last equality we used Eq. \eqref{eq:flm} in conjunction with the fact that $\sum_{\ell m} Y_{\ell m} (\rhat) Y_{\ell m}^* (\rhat^\prime) = \delta^D(\rhat, \rhat^\prime)$.
 
Putting everything together, a properly normalized estimator $\widehat{P}(k)$ of the power spectrum is given by
\begin{equation}
\label{eq:curvedSkyNormFinalResult}
\widehat{P} (k) = \left( \frac{c^2}{2k_B} \right)^2 \frac{|\widetilde{V} (\mathbf{b}, \tau) |^2}{ \int d\Omega d\nu \nu^4 A^2 (\rhat, \nu) \phi^2(r_\nu) \gamma^2 (\nu) / r_\nu^2 \alpha_\nu},
\end{equation} 
where it is understood that the copy of $k$ on the left hand side is tied to the values of $\mathbf{b}$ and $\tau$ on the right hand side via Eq. \eqref{eq:PeakLocs}. Remarkably, this result is almost identical to the estimator previously derived in the literature with many more assumptions (chiefly the flat-sky approximation), reproduced in Appendix \ref{sec:RectilinearInterferometerPspecNorm} for completeness. Comparing Eqs. \eqref{eq:curvedSkyNormFinalResult} and \eqref{eq:flatSkyNormFinalResult}, one sees that the flat-sky approximation has only a minor effect on the result. The two expressions differ only in that with the curved sky case, $r_\nu$ and $\alpha_\nu$ appear inside a radial integral and are evaluated using their full nonlinear expressions, whereas in the flat-sky case, they appear outside the integral and are evaluated at the middle of the radial profile of our survey. Numerically, we find that for the PAPER primary beam, the difference between the Eqs. \eqref{eq:curvedSkyNormFinalResult} and \eqref{eq:flatSkyNormFinalResult} is $\sim 0.1\%$. This rigorously justifies the previous use of flat-sky normalization factors in delay-spectrum-based estimates of the power spectrum \citep{pober_et_al2013b,parsons_et_al2014,ali_et_al2015,jacobs_et_al2015}, regardless of whether an instrument's beam is narrow.

\subsection{The foreground wedge in the spherical Fourier-Bessel formalism}
\label{sec:CurvedSkyWedge}

In Section \ref{sec:WindowFcts}, we saw that our power spectrum window functions became elongated at high $\ell$, providing our first hints of the foreground wedge. However, these hints were not derived in an entirely rigorous fashion, since Section \ref{sec:WindowFcts} and Section \ref{sec:SingleBlNorm} both assumed that the sky temperature is comprised entirely of the cosmological signal. For the purposes of deriving a power spectrum normalization, this is the correct assumption to make. On the other hand, this is insufficient for a derivation of the foreground wedge, since we saw from Section \ref{sec:RotationalInvarianceOnly} that foregrounds have different statistical properties than the cosmological signal.

When the sky consists of more than just the cosmological signal, Eq. \eqref{eq:delayWindow} becomes more complicated because the two-point correlator of $\overline{T}_{\ell m}(k)$ is no longer proportional to the cosmological power spectrum. Instead, foregrounds form an additive contribution to $\overline{T}_{\ell m}(k)$, and---since they are uncorrelated with the cosmological signal---an additive contribution to the two-point correlator. As a simple example, consider the foreground model discussed in Section \ref{sec:RotationalInvarianceOnly}, where the foregrounds possess (statistical) rotation invariance but not translation invariance along the radial/frequency direction. With these foregrounds, Eq. \eqref{eq:delayWindow} becomes
\begin{eqnarray}
\langle |\widetilde{V} (\mathbf{b}, \tau) |^2 \rangle  = \sum_{\ell} \int \! dk \,W_\ell^\textrm{unnorm} (k; \mathbf{b}, \tau) P(k) \nonumber \\
+  \frac{2}{\pi} \sum_{\ell m} C_\ell \Bigg{|} \int \! dk \,k^2 q_\ell (k) g_{\ell m} (k ; \mathbf{b}, \tau) \Bigg{|}^2,
\end{eqnarray}
where $q_\ell (k)$ is the radial spherical Fourier-Bessel transform of the foreground spectrum, as defined by Eq. \eqref{eq:qellk}. Inserting explicit expressions for the $q_\ell$ and $g_{\ell m}$ results in considerable simplifications to the integral in the second term of our expression:
\begin{eqnarray}
&& \int \! dk \, k^2 q_\ell (k) g_{\ell m} (k ; \mathbf{b}, \tau) \nonumber \\
&& = \sqrt{\frac{2}{\pi}} \int \! d\nu dr^\prime e^{i 2\pi \nu \tau}  r^{\prime 2} f_{\ell m} (\mathbf{b}, \nu) \phi(r_\nu) \gamma (\nu)  \kappa(\nu_{r^\prime}) \nonumber \\
&& \quad \times \int dk \,k^2  j_\ell (k r^\prime) j_\ell (k r_\nu) \nonumber \\
&& = \sqrt{\frac{\pi}{2}} \int \! d\nu\, e^{i 2\pi \nu \tau} f_{\ell m} (\mathbf{b}, \nu) \phi(r_\nu) \gamma (\nu) \kappa (\nu),
\end{eqnarray}
where in the second equality we used the orthogonality of spherical Bessel functions from Eq. \eqref{eq:BesselOrthog}. Inserting $f_{\ell m }$ from Eq. \eqref{eq:flm} then gives
\begin{equation}
\sqrt{\frac{\pi}{2}} \int d\Omega B(\rhat) Y_{\ell m} (\rhat) \int \! d\nu\, e^{i 2\pi \nu (\tau - \mathbf{b} \cdot \rhat / c)}  \phi(r_\nu) \gamma (\nu) \kappa (\nu),
\end{equation}
where in this section we are assuming that $B(\rhat)$ is approximately frequency independent in order to highlight the interferometric phenomenology of the foreground wedge. Now, define for notational convenience the quantity $\Theta ( \nu - \nu_0) \equiv \phi(r_\nu) \gamma(\nu_r) \kappa(\nu_r)$ as a re-centered frequency profile of the foregrounds as seen in the data (i.e., including the finite bandwidth $\phi$ of the instrument and the tapering function $\gamma$ imposed by the data analyst). The foreground contribution to $\langle |\widetilde{V} (\mathbf{b}, \tau) |^2 \rangle$ then becomes
\begin{eqnarray}
\label{eq:VtildeSqFgCell}
\langle |\widetilde{V} (\mathbf{b}, \tau) |^2 \rangle \Big{|}_\textrm{fg} = \sum_{\ell m} C_\ell && \Bigg{|} \int d\Omega B(\rhat) Y_{\ell m} e^{i 2\pi \nu_0 (\tau - \mathbf{b} \cdot \rhat / c)} \nonumber \\
&& \times \widetilde{\Theta} \left[ 2 \pi \left( \tau - \frac{\mathbf{b} \cdot \rhat}{c} \right) \right] \Bigg{|}^2.
\end{eqnarray}
To prevent any obscuration of our understanding of the foreground wedge in the spherical Fourier-Bessel formalism, we assume at this point that $C_\ell$ is a constant. As an extreme example of why this is necessary, consider the case where $C_\ell$ is zero everywhere except for one particular value of $\ell$. Clearly, the signature of foregrounds on the $\ell$-$k$ would then be dominated by the rather peculiar form for $C_\ell$, rather than the chromatic interferometric effects we wish to examine here. Setting $C_\ell$ to a constant, our expression reduces to
\begin{eqnarray}
\label{eq:FinalWedgeEq}
\langle |\widetilde{V} (\mathbf{b}, \tau) |^2 \rangle \Big{|}_\textrm{fg} \propto \int d\Omega B^2 (\rhat) \Bigg{|} \widetilde{\Theta} \left[ 2 \pi \left( \tau - \frac{\mathbf{b} \cdot \rhat}{c} \right) \right] \Bigg{|}^2 \nonumber \\
= \frac{c}{b} \int_{2\pi ( \tau - b/c)}^{2\pi ( \tau + b/c)}   ds\overline{B^2} \left[ \arcsin\left( \frac{c \tau}{b} - \frac{sc}{2\pi b} \right) \right] | \widetilde{\Theta} (s) |^2, \quad
\end{eqnarray}
where we performed the polar integral by aligning our polar axis along the direction of the baseline. We then defined $\overline{B^2}$ to be the beam squared profile averaged azimuthally about the baseline axis. However, in the final form of the expression we assumed that the profile has a hemispherical reflection symmetry about the plane perpendicular to the baseline axis, and used this to express $\overline{B^2}$ in a more conventional coordinate system where the polar axis is pointed at zenith.

Eq. \eqref{eq:FinalWedgeEq} contains all the details of the foreground wedge. To make this clear, consider the long baseline limit, which we know from Eq. \eqref{eq:PeakLocs} maps to the high $\ell$ portion of the power spectrum. In this regime, $\overline{B^2}$ is a very broad function of $s$ compared to $\widetilde{\Theta}$, which is compactly localized around $s\approx 0$ (since $\Theta$ is a centered spectral profile) for spectrally smooth foregrounds that are surveyed by an instrument with broad frequency coverage. We may thus factor $\overline{B^2}$ out of the integral, evaluating it at $s=0$ to give
\begin{equation}
\label{eq:approxFinalWedgeEq}
\langle |\widetilde{V} (\mathbf{b}, \tau) |^2 \rangle \Big{|}_\textrm{fg} \appropto \frac{c}{b} \overline{B^2} \left[ \arcsin\left( \frac{c \tau}{b}  \right) \right] \int_{2\pi ( \tau - b/c)}^{2\pi ( \tau + b/c)}   ds| \widetilde{\Theta} (s) |^2.
\end{equation}
There are two key features to this equation. The first is that $\langle |\widetilde{V} (\mathbf{b}, \tau) |^2 \rangle$ is zero if $\tau$ is not within $\pm b / c$ of zero, because $\widetilde{\Theta}$ is peaked around zero. This means that there will be no foreground emission beyond $\tau > b/c$. Inserting Eq. \eqref{eq:PeakLocs} into this condition implies that foreground on the $\ell$-$k$ plane will be restricted to
\begin{equation}
k < \ell \left( \frac{1}{\alpha_0^2 \nu_0^2}+\frac{1}{r_0^2}\right)^{\frac{1}{2}},
\end{equation}
or in terms cosmological quantities,
\begin{equation}
\label{eq:Finalkellwedge}
k <  \ell \frac{H_0}{c} \left[\frac{E^2(z)}{(1+z)^2} + \left( \int_0^z \frac{dz^\prime}{E(z^\prime)}\right)^{-2}\right]^{\frac{1}{2}}.
\end{equation}
We therefore have a wedge signature (beyond which there is minimal foreground contamination) similar to what is seen on the $k_\perp$-$k_\parallel$ plane. This is seen in the bottom right panel of Figure \ref{fig:fgSigs}, where we numerically evaluate Eq. \eqref{eq:VtildeSqFgCell} for a flat intrinsic angular power spectrum for the foregrounds, with survey parameters kept the same as they were in previous sections.

The other key feature Eq. \eqref{eq:approxFinalWedgeEq} is the way in which foreground power drops off as one approaches the edge of the wedge. For regions of the $\ell$-$k$ plane that satisfy Eq. \eqref{eq:Finalkellwedge} (i.e., ``inside/below the wedge"), the integral in Eq. \eqref{eq:approxFinalWedgeEq} evaluates to a constant factor, leaving a spherical harmonic power spectrum signature $\widehat{S}_\ell^\textrm{fg} (k)$ of the form\footnote{Note that while our results for the boundary of the wedge and its profile are qualitatively robust, minor differences can arise depending on the precise form of the power spectrum estimator that is employed. Consider, for example, the estimator used in \citet{liu_et_al2014a} where visibility data was convolved onto a Fourier-space grid using the primary beam as a gridding kernel. There, the profile of the wedge was shown to be primary beam convolved with itself, rather than the primary beam squared as we have it here for our estimator.}
\begin{eqnarray}
\label{eq:Sellfgk}
\widehat{S}_\ell^\textrm{fg} (k) &\propto& \langle |\widetilde{V} (\mathbf{b}, \tau) |^2 \rangle \Big{|}_\textrm{fg}  \nonumber \\
&\appropto&\frac{1}{\ell} \overline{B^2} \left[ \arcsin\left( \alpha_0 \nu_0 \sqrt{\frac{k^2}{\ell^2} - \frac{1}{r_0^2}}  \right) \right].
\end{eqnarray}
Ignoring the $1/\ell$ prefactor (which only weakly tilts the power profile), this expression shows that for regions within the wedge on the $k$-$\ell$ plane, contours of constant power take the form of straight lines where $k \propto \ell$. As $k$ increases, these contours decrease in power with a profile determined by the square of the beam, averaged along the direction perpendicular to the baseline.

Eq. \eqref{eq:Sellfgk} does not hold for short baselines (i.e., at low $\ell$) because the approximations that led to Eq. \eqref{eq:approxFinalWedgeEq} no longer apply. In such a regime, the profile becomes proportional to $| \widetilde{\Theta} ( \alpha_0 k )|^2$, leading to the horizontally oriented power patterns seen at low $\ell$ in Figure \ref{fig:fgSigs}. This contrast in behavior between low and high $\ell$ regions is familiar from the $k_\perp$-$k_\parallel$ plane: at low $\ell$ or low $k_\perp$, the leakage of flat-spectrum foregrounds towards the upper portions of the plane is driven by the radial extent of the survey, while at high $\ell$ or high $k_\perp$ the leakage is driven by the baseline chromaticity that causes the wedge. In intermediate regimes, Eq. \eqref{eq:FinalWedgeEq} is similar in form to a convolution. In fact, it would be precisely a convolution were it not for the $\arcsin$ and the some constant factors needed for unit conversions. This convolution-like operation enacts a smooth transition in behavior between the low- and high-$\ell$ regimes.

Fundamentally, the wedge signature arises because the chromaticity of an interferometer causes spectrally smooth foregrounds from low $k$ or $k_\parallel$ (as seen in the top row of Figure \ref{fig:fgSigs}) to leak to higher $k$ and $k_\parallel$. In other words, power is smeared out along the $k$ or $k_\parallel$ axes. Though its most dominant effect is to cause the foreground wedge, this smearing affects all modes on the $k_\perp$-$k_\parallel$ and $\ell$-$k$ planes, particularly at high $k_\perp$ and high $\ell$ where chromatic effects are more prominent. This can be seen by examining Figure \ref{fig:windowFcts}, where the window functions for the cosmological signal are seen to vertically broaden at high $\ell$, regardless of location along the $k$ axis. (In principle, Figure \ref{fig:windowFcts} only applies to signals that possess translation-invariant statistics, but the effects are qualitatively the same). The broadening with increasing $\ell$ can be seen by comparing the non-interferometric (top row) and interferometric (bottom row) results in Figure \ref{fig:fgSigs}. As discussed in Section \ref{sec:Numerics}, the cosine radial profile given by Eq. \eqref{eq:CosineRadial} causes ringing in Fourier space that gives horizontal stripes that are visually obvious in the non-interferometric case. For the interferometric case, the ringing is still present, but the peaks are smeared out, especially at high $\ell$. This reinforces what was found in \citet{liu_et_al2014a}, where it was argued that chromatic interferometric effects do not only cause the wedge, but also reduce the independence of different Fourier modes.

In summary, we have seen in this section that the spherical power spectrum provides the same foreground diagnostic capabilities on the $\ell$-$k$ plane as the rectlinear power spectrum did on the $k_\perp$-$k_\parallel$ plane. In the spherical Fourier-Bessel formalism, the foregrounds continue to be confined to a wedge. This is good news for analysts of wide-field intensity mapping data from interferometers, for it suggests that one's intuition for the $k_\perp$-$k_\parallel$ plane can be easily transferred to the $\ell$-$k$ plane.

\section{Conclusions}
\label{sec:Conclusions}

In this paper, we have established a framework for analyzing intensity mapping data using spherical Fourier-Bessel techniques. Such techniques easily incorporate the wide-field nature of many intensity mapping surveys, obviating the need to split up one's field into several approximately flat fields during analysis. This builds sensitivity for science measurements as well as diagnostic tests, and additionally provides access to the largest angular scales on the sky.

Adapting spherical Fourier-Bessel techniques from galaxy surveys requires one to pay special attention to the unique properties of intensity mapping. For example, we saw in Figure \ref{fig:surveyGeom} that intensity mapping surveys (particularly those that operate at high redshifts) tend to be compressed in the radial direction and have very fine radial resolution compared to angular resolution. Intensity mapping experiments must also contend with extremely bright foregrounds that overwhelm the cosmological signals of interest. A successful spherical Fourier-Bessel analysis framework must demonstrate that it is able to deal with such systematics at least as well as traditional rectilinear Fourier techniques can.

This paper demonstrates that spherical Fourier-Bessel modes are indeed a suitable basis for intensity mapping analyses. Focusing on power spectrum measurements, in Section \ref{sec:FiniteVolume} we proposed that the cylindrically binned power spectrum $P(k_\perp, k_\parallel)$ be replaced by the spherical harmonic power spectrum $S_\ell (k)$. The quantity $S_\ell (k)$ is conveniently defined so that a weighted average of it over different $\ell$ values yields the spherically binned cosmological power spectrum $P(k)$. At the same time, by splitting up the measured power spectrum into a function of $\ell$ and $k$, angular fluctuations are separated from arbitrarily oriented spatial fluctuations. This separation of fluctuations into angular and non-angular modes provides a powerful diagnostic for systematics. This has historically been the motivation for viewing the power spectrum as a function of $k_\perp$ and $k_\parallel$, and $S_\ell (k)$ preserves this crucial property of $P(k_\perp, k_\parallel)$. Of course, this is not to say that the data should not also be examined in bases like $(\mathbf{b}, \tau)$ that are more closely related to the actual instrument's measurement \citep{Vedantham2012,Parsons_et_al2012b}. Doing so is particularly valuable prior to the squaring of the data to form power spectra, and both approaches can and should be used.

Chief amongst the systematics that may be discerningly diagnosed on the $k_\perp$-$k_\parallel$ plane are astrophysical foregrounds. Foregrounds are expected to have localized signatures in $P(k_\perp, k_\parallel)$, facilitating their removal. We have shown in this paper that the same is true for $S_\ell (k)$. For non-interferometric intensity mapping surveys, we have shown that the spectrally smooth nature of foregrounds results in their being sequestered at low $k$, and that interloper lines can be detected using $S_\ell (k)$ just as easily as they can be using $P(k_\perp, k_\parallel)$. For interferometric surveys, foregrounds tend to limited to a wedge-like feature on the $k_\perp$-$k_\parallel$ plane. Foregrounds are limited to a similar wedge on the $\ell$-$k$ plane. This suggests that $S_\ell(k)$ is just as capable a diagnostic quantity as $P(k_\perp, k_\parallel)$ for intensity mapping surveys, while simultaneously discarding unwarranted flat-sky approximations seen in previous papers. Another attractive property of our spherical Fourier-Bessel formulation is that many of the relevant formulae derived in this paper (such as the equation delineating the boundary of the foreground wedge) are very similar to their flat-sky counterparts. Intuition for the behavior of $P(k_\perp, k_\parallel)$ that has been built up in the prior literature is thus almost entirely transferrable to $S_\ell (k)$.

Our framework may be generalized in several ways in future work. For instance, we have thus far neglected to describe redshift space distortions, although the spherical formalism that we espouse here should be particularly well-suited for the purpose (C. J. Schmit et al., in prep.). A crucial area of investigation will be to determine whether cosmological redshift space distortions interfere with the signature of interloper lines. Another area of future development would be the incorporation of light-cone effects, since it has been shown that cosmological evolution cannot be neglected over the survey volume of a typical intensity mapping survey \citep{barkana_and_loeb2006,datta_et_al2012,datta_et_al2014,laplante_et_al2014,zawada_et_al2014,ghara_et_al2015}. For now, however, this paper points to the promise of spherical Fourier-Bessel techniques for rigorous data analysis, providing yet another powerful diagnostic tool in the continuing progress of intensity mapping towards surveying an unprecedentedly large volume of our observable Universe.

\acknowledgments

The authors gratefully acknowledge delightful and helpful discussions with James Aguirre, Michael Eastwood, Aaron Ewall-Wice, Daniel Jacobs, Gregg Hallinan, Bryna Hazelton, Jacqueline Hewitt, Saul Kohn, Miguel Morales, Jonathan Pober, Jonathan Pritchard, Claude Schmit, Richard Shaw, and Nithya Thyagarajan. This research was completed as part of the University of California Cosmic Dawn Initiative. AL and ARP acknowledge support from the University of California Office of the President Multicampus Research Programs and Initiatives through award MR-15-328388, as well as from NSF CAREER award No. 1352519, NSF AST grant No.1129258, and NSF AST grant No. 1440343. AL acknowledges support for this work by NASA through Hubble Fellowship grant \#HST-HF2-51363.001-A awarded by the Space Telescope Science Institute, which is operated by the Association of Universities for Research in Astronomy, Inc., for NASA, under contract NAS5-26555. This research used resources of the National Energy Research Scientific Computing Center, a DOE Office of Science User Facility supported by the Office of Science of the U.S. Department of Energy under Contract No. DE-AC02-05CH11231.

\appendix
\section{Estimating the power spectrum from rectilinear Fourier modes}
\label{sec:RectilinearFKP}

In this Appendix, we derive a relation between measured rectilinear Fourier amplitudes of the sky $\widetilde{T}^\textrm{meas} (k)$ and the power spectrum, analogous to Eq. \eqref{eq:TlmPkProportionality} for the spherical Fourier-Bessel modes. The derivation presented here is a standard one, and is only included to serve as an analogy to Eq. \eqref{eq:TlmPkProportionality}.

For a survey specified by the function $\phi (\mathbf{r})$---so that the measured temperature field is $\phi(r) T(\mathbf{r})$ rather than just $T(\mathbf{r})$---the measured Fourier amplitudes $\widetilde{T}^\textrm{meas} (\mathbf{k})$ are related to the true Fourier amplitudes $\widetilde{T} (\mathbf{k})$ by the convolution theorem, which gives
\begin{equation}
\widetilde{T}^\textrm{meas} (\mathbf{k}) = \int \frac{d^3 k^\prime}{(2 \pi)^3} \widetilde{T} (\mathbf{k}^\prime) \widetilde{\phi} (\mathbf{k} - \mathbf{k}^\prime).
\end{equation} 
Squaring and ensemble averaging the result gives
\begin{equation}
\langle | \widetilde{T}^\textrm{meas} (\mathbf{k}) |^2 \rangle = \int \frac{d^3 k^\prime}{(2 \pi)^3} \big{|} \widetilde{\phi} (\mathbf{k} - \mathbf{k}^\prime) \big{|}^2 P(k^\prime),
\end{equation}
where we used Eq. \eqref{eq:RectilinearPspecDef} to relate the ensemble average of the true Fourier amplitudes to the power spectrum. Assuming $|\widetilde{\phi}|^2$ is sharply peaked, $P(k^\prime)$ can be approximately factored out of the integral and evaluated at $k$. Changing integration variables from $\mathbf{k}^\prime$ to $\mathbf{k} - \mathbf{k}^\prime$ for the remaining integral and invoking Parseval's theorem then yields
\begin{equation}
\langle | \widetilde{T}^\textrm{meas} (\mathbf{k}) |^2 \rangle = P(k) \int d^3 r \phi^2 (\mathbf{r}).
\end{equation}
This suggests that a power spectrum estimator $\widehat{P} (k)$ can be constructed by computing
\begin{equation}
\widehat{P} (k) = \frac{\sum_{|\mathbf{k}| = k} | \widetilde{T}^\textrm{meas} (\mathbf{k}) |^2}{N_k \int d^3 r \phi^2 (\mathbf{r})},
\end{equation}
where $N_k$ is the number of independent Fourier modes in the shell where $|\mathbf{k}| = k$. The rest of the normalization factor that comprises the denominator (i.e., the integral) is independent of $k$ and is a sensitivity-weighted volume factor. For a survey with uniform sensitivity, for example, $\phi(\mathbf{r}) = 1$ everywhere inside the survey and the integral is exactly the volume of the survey. Because this integral is independent of $\mathbf{k}$, it follows that the orientation of a Fourier mode (i.e., $\khat$) has no bearing on its sensitivity to the power spectrum, and all orientations are equally sensitive.

\section{Delay spectrum normalization in the narrow-field flat-sky limit}
\label{sec:RectilinearInterferometerPspecNorm}

For orientation, we now briefly review how visibility-based estimators of the power spectrum are usually derived. Again, the derivation that follows is not new to this paper, but we include it to provide a pedagogical comparison to Section \ref{sec:SingleBlNorm}, as well as to place a special emphasis on the approximations involved.

Our first approximation will be the flat-sky, narrow-field approximation. This makes Eq. \eqref{eq:FlatVisibility} the appropriate expression to use for our interferometric visibility. Next, we assume that the interferometric fringe in this visibility is frequency-independent, so that the factor of $\nu$ in the complex exponential term may be replaced by $\nu_0$, the median frequency of one's observing volume. This is equivalent to the approximation that one has very short baselines (since the baseline vector $\mathbf{b}$ is multiplied by $\nu$), or alternatively, that most spectral structure comes from either the primary beam or the sky temperature. Correspondingly, we also replace all copies of $r_\nu$ with $r_0$ to yield\footnote{In principle, converting $\nu_0$ to a radial distance does not yield $r_0$ because the distance-frequency relation is nonlinear. In practice, the radially compressed geometry of a typical intensity mapping survey (see Figure \ref{fig:surveyGeom}) means that linearized distance-frequency relations such as Eq. \eqref{eq:LinearDistanceApprox} are excellent approximations. We are thus justified in making the assumption that $\nu_0$ and $r_0$ roughly refer to the same radial distance.}
\begin{equation}
V(\mathbf{b}, \nu) = \frac{1}{r_0^2}\int d^2 r_\perp  \phi(r) B(\mathbf{r_\perp}, \nu) T(\mathbf{r_\perp}, \nu)e^{ - i 2\pi \nu_0  \mathbf{b} \cdot \mathbf{r}_\perp / c r_0}.
\end{equation}
To access Fourier modes along the line-of-sight, we perform the delay transform defined by Eq. \eqref{eq:DelayDef}. Converting again to cosmological coordinates and defining
\begin{equation}
\label{eq:kperpConversions}
\mathbf{k}_\perp \equiv \frac{2 \pi \nu_0}{r_0 c} \mathbf{b}; \qquad k_\parallel \equiv \frac{2 \pi \tau}{\alpha},
\end{equation}
along with $\mathbf{k} \equiv (\mathbf{k}_\perp, k_\parallel)$, one obtains
\begin{eqnarray}
\widetilde{V}(\mathbf{b}, \tau) & = &  \frac{e^{i 2\pi \nu_0  \tau - i k_\parallel r_0}}{r_0^2 \alpha_0 } \int d^3r D(\mathbf{r})T(\mathbf{r}) \exp \left( - i \mathbf{k} \cdot \mathbf{r} \right) \nonumber \\
& = &  \frac{e^{i 2\pi \nu_0  \tau - i k_\parallel r_0}}{r_0^2 \alpha_0 }  \int \frac{d^3 k^\prime}{(2 \pi)^3} \widetilde{T} (\mathbf{k}^\prime) \widetilde{D} ( \mathbf{k} - \mathbf{k}^\prime),
\end{eqnarray}
where $\gamma$ is the tapering functions used in Eq. \eqref{eq:DelayDef}, $\alpha_0$ is the distance-frequency conversion from Eq. \eqref{eq:AlphaConversion} evaluated at the redshift corresponding to the central frequency of the survey, and we have defined $D(\mathbf{r}) \equiv B(\mathbf{r}) \phi(r) \gamma(\nu_{r_\parallel}) $, with $\widetilde{D}$ denoting its Fourier transform. While survey geometry and tapering factors such as $\phi$ and $\gamma$ have not typically been included in literature derivations such as those in \citet{parsons_et_al2012a} and \citet{parsons_et_al2014}, they are crucial in practical analyses of the data (e.g., in \citealt{ali_et_al2015}), and thus we include them here.

As suggested in Section \ref{sec:DelayIntro}, we may relate the delay-transformed visibility to the power spectrum by squaring $\widetilde{V}(\mathbf{b}, \tau)$ and taking an ensemble average over realizations of the random temperature field. This gives
\begin{eqnarray}
\langle | \widetilde{V}(\mathbf{b}, \tau) |^2 \rangle &= & \left( \frac{1}{r_0^2 \alpha_0 } \right)^2 \int \frac{d^3 k^\prime}{(2 \pi)^3} \frac{d^3 k^{\prime \prime}}{(2 \pi)^3} \nonumber \\
&& \widetilde{D} ( \mathbf{k} - \mathbf{k}^\prime) \widetilde{D}^* ( \mathbf{k} - \mathbf{k}^{\prime \prime}) \langle \widetilde{T} (\mathbf{k}^\prime) \widetilde{T}^* (\mathbf{k}^{\prime \prime}) \rangle \nonumber \\
& = & \left( \frac{1}{r_0^2 \alpha_0 } \right)^2 \int \frac{d^3 k^\prime}{(2 \pi)^3} P(\mathbf{k}^\prime) |\widetilde{D} ( \mathbf{k} - \mathbf{k}^{\prime})|^2, \qquad
\end{eqnarray}
where in the last equality we invoked the definition of the power spectrum, i.e., Eq. \eqref{eq:RectilinearPspecDef}. At this point, we may make the approximation that the power spectrum is a relatively broad function, while $|\widetilde{D} ( \mathbf{k} - \mathbf{k}^{\prime})|^2$ is fairly sharply peaked at $\mathbf{k} = \mathbf{k}^\prime$. This allows $P (\mathbf{k})$ to be factored out of the integral, and by invoking Parseval's theorem on what remains, we obtain
\begin{equation}
\langle | \widetilde{V}(\mathbf{b}, \tau) |^2 \rangle \approx \left( \frac{1}{r_0^2 \alpha_0 } \right)^2 P (\mathbf{k}) \int d^3 r D^2 (\mathbf{r}).
\end{equation}
A sensible estimator $\widehat{P} (\mathbf{k})$ for the power spectrum would thus be
\begin{eqnarray}
\widehat{P} (\mathbf{k}) &=& r_0^4 \alpha^2 \left[ \int d^3 r D^2 (\mathbf{r}) \right]^{-1} | \widetilde{V}(\mathbf{b}, \tau) |^2 \nonumber \\
& =&  \frac{\alpha  r_0^2 }{ \int d^2\theta d\nu B^2 (\btheta, \nu) \gamma(\nu) \phi(r_\nu)}  | \widetilde{V}(\mathbf{b}, \tau) |^2. \qquad\quad
\end{eqnarray}
Now, even though this expression was derived using the flat-sky approximation, it has been applied to wide-field instruments in the past. The flat-sky approximation is crudely undone by 
promoting $d^2 \theta$ back to $d \Omega$, giving
\begin{equation}
\label{eq:flatSkyNormFinalResult}
\widehat{P} (\mathbf{k}) = \left( \frac{ c^2}{2 k_B } \right)^2 \frac{r_0^2 \alpha  | \widetilde{V}(\mathbf{b}, \tau) |^2 }{\int \!d\nu d\Omega \,\nu^4 A^2 (\rhat, \nu)\gamma(\nu) \phi(r_\nu) },
\end{equation}
where we have reinserted Eq. \eqref{eq:Bdef}. It is implicitly assumed that the value of $\mathbf{k}$ on the left hand side of this equation is related to $\mathbf{b}$ and $\tau$ by Eq. \eqref{eq:kperpConversions}.

\section{The foreground wedge in the narrow-field limit}

In this Appendix, we work in the narrow-field limit and derive an analytic form for the signature of foregrounds in a power spectrum expressed in terms of rectilinear $k_\perp$-$k_\parallel$ Fourier modes (i.e., the ``foreground wedge" on the $k_\perp$-$k_\parallel$ plane). Our starting point will be Eq. \eqref{eq:FlatVisibility}, but written in terms of angles on the sky and assuming a frequency-independent modified primary beam $B(\boldsymbol \theta)$:
\begin{equation}
V(\mathbf{b}, \nu) = \int d^2 \theta \phi(r_\nu) B(\boldsymbol \theta) T(\boldsymbol \theta, \nu)e^{ - i 2\pi \nu  \mathbf{b} \cdot \boldsymbol \theta / c}.
\end{equation}
The delay-transformed visibility then takes the form
\begin{equation}
\widetilde{V} (\mathbf{b}, \tau) = \int d\nu d^2 \theta \gamma(\nu) \phi(r_\nu) B(\boldsymbol \theta) T(\boldsymbol \theta, \nu) e^{i 2 \pi \nu ( \tau - \mathbf{b} \cdot \boldsymbol \theta / c)}.
\end{equation}

In principle, our sky temperature $T(\boldsymbol \theta, \nu)$ should include contributions from both the cosmological signal and the foregrounds. However, if we assume that foregrounds and the cosmological signal are uncorrelated (as we did in Section \ref{sec:CurvedSkyWedge}), we may derive the foreground wedge without including the cosmological signal. We thus assume in this Appendix that $T(\boldsymbol \theta, \nu)$ consists solely of foregrounds, taking the form 
\begin{equation}
T(\boldsymbol \theta, \nu) = T_\perp^\textrm{fg} (\boldsymbol \theta) \kappa(\nu),
\end{equation}
and thus our delay-transformed visibility becomes
\begin{eqnarray}
\widetilde{V} (\mathbf{b}, \tau) & = & \int d^2 \theta B(\boldsymbol \theta) T_\perp^\textrm{fg} (\boldsymbol \theta) e^{i 2 \pi \nu_0 ( \tau - \mathbf{b} \cdot \boldsymbol \theta / c)} \nonumber \\
&& \qquad \times \widetilde{\Theta} \left[ 2 \pi \left( \tau - \frac{\mathbf{b} \cdot \boldsymbol \theta}{c}\right)\right]  \nonumber \\
&=& \int \frac{d^2 \ell}{(2\pi)^2} \widetilde{T}_\perp^\textrm{fg} (\boldsymbol \ell) \int d^2 \theta B(\boldsymbol \theta) e^{i \boldsymbol \ell \cdot \boldsymbol \theta} e^{i 2 \pi \nu_0 ( \tau - \mathbf{b} \cdot \boldsymbol \theta / c)} \nonumber \\
&& \qquad \times \widetilde{\Theta} \left[ 2 \pi \left( \tau - \frac{\mathbf{b} \cdot \boldsymbol \theta}{c}\right)\right],
\end{eqnarray}
where $\widetilde{T}_\perp^\textrm{fg} (\boldsymbol \ell) \equiv \int d^2\theta e^{i \boldsymbol \ell \cdot \boldsymbol \theta} T_\perp^\textrm{fg} (\boldsymbol \theta)$ is the Fourier transform of $T_\perp^\textrm{fg}$, which we assume (as we did in Sections \ref{sec:RotationalInvarianceOnly} and \ref{sec:CurvedSkyWedge}) is a field with rotationally invariant statistics. This means that
\begin{equation}
\langle \widetilde{T}_\perp^\textrm{fg} (\boldsymbol \ell) \widetilde{T}_\perp^\textrm{fg} (\boldsymbol \ell)^* \rangle = (2\pi)^2 \delta^D (\boldsymbol \ell - \boldsymbol \ell^\prime) C_\ell^\textrm{fg},
\end{equation}
where we have suggestively chosen the symbol $C_\ell^\textrm{fg}$ on the right hand side because in the flat-sky limit, $C_\ell^\textrm{fg}$ can be shown to converge to the angular power spectrum \citep{hu2000}.

Following Section \ref{sec:CurvedSkyWedge}, we form our estimator of the power spectrum by squaring the absolute magnitude of the delay-transformed visibility to obtain
\begin{eqnarray}
\langle | \widetilde{V} (\mathbf{b}, \tau) |^2 \rangle = \int \frac{d^2 \ell}{(2\pi)^2} C_\ell^\textrm{fg} \Bigg{|} \int d^2 \theta e^{i (\boldsymbol \ell - 2 \pi \nu_0 \mathbf{b} / c) \cdot \boldsymbol \theta} \nonumber \\
\times B(\boldsymbol \theta) \widetilde{\Theta} \left[ 2 \pi \left( \tau - \frac{\mathbf{b} \cdot \boldsymbol \theta}{c}\right)\right] \Bigg{|}^2. \qquad
\end{eqnarray}
Again, we may consider the special case where $C_\ell^\textrm{fg}$ is a constant in order to elucidate the effects of the foreground wedge. This yields
\begin{eqnarray}
\langle | \widetilde{V} (\mathbf{b}, \tau) |^2 \rangle &\propto& \int d^2 \theta B^2 (\boldsymbol \theta) \Bigg{|} \widetilde{\Theta} \left[ 2 \pi \left( \tau - \frac{\mathbf{b} \cdot \boldsymbol \theta}{c} \right) \right] \Bigg{|}^2 \nonumber \\
&=& 2 \pi \int d \theta \overline{B^2} (\theta) \widetilde{\Theta} \left[ 2 \pi \left( \tau - \frac{b \theta}{c} \right) \right] \Bigg{|}^2 \nonumber \\
&=& \frac{c}{b} \int ds \overline{B^2} \left( \frac{c \tau}{b} - \frac{sc}{2\pi b} \right) | \widetilde{\Theta} (s) |^2 \nonumber \\
&\approx & \frac{c}{b} \overline{B^2} \left( \frac{c \tau}{b} \right) \int ds | \widetilde{\Theta} (s) |^2
\end{eqnarray}
where $\overline{B^2} (\theta) \equiv (1/2\pi) \int d\theta^\prime B^2(\theta, \theta^\prime) $, and in the last line we assumed we were in the long baseline (or the high $k_\perp$) regime where the foreground wedge is relevant. This allowed $\overline{B^2}$ to be factored out of the integral.

Recalling that $\langle | \widetilde{V} (\mathbf{b}, \tau) |^2 $ serves as a good estimator for the power spectrum at Fourier coordinates given by Eq. \eqref{eq:kperpConversions}, the foreground contamination $\widehat{P}^\textrm{fg}$ of our power spectrum estimate is thus given by
\begin{equation}
\widehat{P}^\textrm{fg} (k_\perp, k_\parallel) \propto \frac{1}{k_\perp}  \overline{B^2} \left( \frac{\alpha_0 \nu_0 k_\parallel}{r_0 k_\perp}\right).
\vspace{0.1cm}
\end{equation}
Contours of constant foreground power are therefore along lines where $k_\parallel \propto k_\perp$, and if $\overline{B^2}$ is zero (or negligible) beyond some characteristic angle $\theta_c$ away from its peak, foreground emission will be confined to
\begin{equation}
\label{eq:WedgeLineEquation}
k_\parallel < k_\perp \frac{H_0 r_0 E(z) \theta_c}{c (1+z)} ,
\end{equation}
where we have written $\alpha_0$ in terms of cosmological parameters. Since $k = (k_\perp^2 + k_\parallel^2)^{1/2}$, we may also write this in terms of $k$ and $k_\perp$. This gives
\begin{equation}
k < k_\perp r_0 \frac{H_0}{c} \left[\frac{\theta_c^2 E^2(z)}{(1+z)^2} + \left( \int_0^z \frac{dz^\prime}{E(z^\prime)}\right)^{-2}\right]^{\frac{1}{2}}.
\end{equation}
\\
Recalling that $\ell \approx k_\perp r_0$ in the flat-sky approximation, this is essentially the same as the full curved-sky expression, Eq. \eqref{eq:Finalkellwedge}. The only slight difference is that in the flat-sky approximation, the angular coordinates are rectilinear and formally go from $-\infty$ to $+\infty$, necessitating some arbitrary cut-off angle $\theta_c$ for the primary beam. In the curved sky treatment, a cutoff is naturally imposed by the horizon.


\bibliographystyle{apj}
\bibliography{biblio}

\end{document}